\documentclass[fleqn,1p]{elsarticle}
\usepackage{fullpage}
\usepackage{microtype, natbib} 
\usepackage[lined,ruled,vlined]{algorithm2e}
\usepackage{graphicx, booktabs}
\graphicspath{{figures/}}
\usepackage{subfigure}
\usepackage{booktabs} %
\usepackage{amsfonts, amssymb, amsthm, amsmath, hyperref, cleveref,chemformula}
\usepackage[export]{adjustbox} 
\usepackage{tcolorbox}
\tcbuselibrary{theorems}
\usepackage{xparse}
\usepackage[textwidth=60pt]{todonotes}

\usepackage{tikz,array}
\usetikzlibrary{shapes.geometric}
\usetikzlibrary{shapes.arrows}

\usepackage{forest}

\newcommand{\dr}{\mathrm{d} \mathbf{r}}

\definecolor{myred}{rgb}{0.7,0.1,0.45}
\definecolor{myrvs}{rgb}{0.25,0.45,0.85}
\definecolor{whucolor}{rgb}{0.29,0.33,0.13}
\definecolor{um0color}{rgb}{1.1,0.01,0.24}
\definecolor{umcolor}{rgb}{1.0,0.01,0.24}
\definecolor{nuscolor}{rgb}{0.0,0.3,0.62}

\hypersetup{
	colorlinks=true,
	linkcolor=nuscolor,
	citecolor=nuscolor,
	filecolor=magenta,      
	urlcolor=cyan,
	pdftitle={HarMul},
	pdfpagemode=FullScreen,
}

\title{Towards chemical accuracy using a multi-mesh adaptive finite element method in all-electron density functional theory}

\author[1]{Yang Kuang}
\ead{ykuang@gdut.edu.cn}
\author[2,3]{Yedan Shen}
\author[4,5]{Guanghui Hu\corref{cor1}}
\cortext[cor1]{Corresponding author. Email:{\tt garyhu@um.edu.mo}}

\address[1]{School of Mathematics and Statistics \&  Center for Mathematics and Interdisciplinary Science (CMIS), Guangdong University of Technology,  China}
\address[2]{School of Mathematics and Information Science, Guangzhou University, China}
\address[3]{Hunan Key Laboratory for Computation and Simulation in Science and Engineering, Xiangtan University, China.}
\address[4]{Department of Mathematics \& Guangdong-Hong Kong-Macao Joint Laboratory for Data-Driven Fluid Mechanics and En- gineering Applications, University of Macau, China }
\address[5]{Zhuhai UM Science and Technology Research Institute, Zhuhai, China.}
\date{\today}

\begin{document}
	\begin{abstract}
		Chemical accuracy serves as an important metric for assessing the effectiveness of the numerical method in Kohn--Sham density functional theory. It is found that to achieve chemical accuracy, not only the Kohn--Sham wavefunctions but also the Hartree potential, should be approximated accurately. Under the adaptive finite element framework, this can be implemented by constructing the \emph{a posteriori} error indicator based on approximations of the aforementioned two quantities. However, this way results in a large amount of computational cost. To reduce the computational cost, we propose a novel multi-mesh adaptive method, in which the Kohn--Sham equation and the Poisson equation are solved in two different meshes on the same computational domain, respectively. With the proposed method, chemical accuracy can be achieved with less computational consumption compared with the adaptive method on a single mesh, as demonstrated in a number of numerical experiments.
	\end{abstract}
	\begin{keyword}
		Chemical accuracy; Kohn--Sham equation;  Adaptive finite element method; Multi-mesh method
	\end{keyword}
	
	\maketitle
	
	\section{Introduction}
	\label{sec:intro}
	The Kohn--Sham density functional theory plays an important role in the area of quantum physics, condensed matter, and computational chemistry. For a quantum system containing $N_{nuc}$ nuclei and $N_{ele}$ electrons, the ground state can be obtained by solving the lowest $N_{occ}$ eigenpairs ${(\varepsilon_i,\psi_i)}_{i=1,\dots,N_{occ}}$ from the Kohn--Sham equation  
	\begin{equation}\label{eq:ks}
		\left(-\frac{1}{2}\nabla^2+V_\mathrm{ext}(\mathbf{r})+V_\mathrm{Har}(\mathbf{r})+V_\mathrm{xc}(\mathbf{r})\right)\psi_i(\mathbf{r}) = \varepsilon_i\psi_i(\mathbf{r}),~i=1,\dots,N_{occ},
	\end{equation}
	where $N_{occ}=N_{ele}/2$ if $N_{ele}$ is even and $(N_{ele}+1)/2$ if $N_{ele}$ is odd is the number of occupation number. $V_\mathrm{ext}(\mathbf{r}) = -\sum_I Z_I/|\mathbf{R}_I-\mathbf{r}|$ represents the external potential due to the nuclei located at $\{\mathbf{R}_I\}_{I=1,\dots,N_{nuc}}$ with charge $\{Z_I\}_{I=1,\dots,N_{nuc}}$. $V_\mathrm{Har}(\mathbf{r})$ stands for the Hartree potential which describes the interaction among electrons. $V_\mathrm{xc}(\mathbf{r})$ is the exchange-correlation potential used to absorb the difference between the Kohn--Sham one-body non-interactive system and the real many-body interactive system.
	
	Various numerical methods have been presented to solve the Kohn-Sham density functional theory \cite{wimmer1981full, ahlrichs1989electronic, yang2007trust, bao2012h, cohen2013locally, motamarri2013higher, schauer2013all, davydov2016adaptive}. To verify the effectiveness of these methods, the ability to achieve chemical accuracy has been a commonly accepted criterion. The chemical accuracy is generally considered as 1 kcal/mole accuracy ($\approx 1.59\times 10^{-3}$ Hartree/particle) with respect to the total energy \cite{pople1999nobel}. In this work, we consider the chemical accuracy as the total energy to achieve the accuracy within $1\times 10^{-3}$ Hartree/atom compared with the energy obtained from reference software.
	
	Due to the rapid variations of the wavefunctions around the nuclear positions, a uniform discretization of the Kohn--Sham equation in the computational domain would result in an unaffordable implementation, especially in all-electron calculations. The adaptive finite element (AFE) methods which enable different mesh sizes are favored in solving the Kohn--Sham equation. Furthermore, the ability of handling the non-periodic boundary condition and complex computational domain makes AFE methods more and more attractive in electronic structure calculation in recent decades \cite{tsuchida1995electronic, bao2012h, motamarri2013higher,chen2014adaptive, maday2014h, davydov2016adaptive, motamarri2020dft}. 
	
	There have been lots of efforts for efficiently achieving chemical accuracy by the adaptive finite element methods, such as using the high-order finite elements \cite{lehtovaara2009all, motamarri2013higher, schauer2013all, maday2014h, davydov2016adaptive, fang2012kohn}, high-order numerical quadrature \cite{schauer2013all, davydov2016adaptive}, etc. In addition, an accurate approximation to the Hartree potential is shown to be key in achieving chemical accuracy \cite{tsuchida1995electronic, motamarri2013higher, schauer2013all, davydov2016adaptive}. Under the finite element discretization framework, the Hartree potential is generally obtained from solving a Poisson equation on the same computational domain as the Kohn--Sham equation. It is noted that the decay of Hartree potential in the infinity behaves as $1/r$ which is much slower than the exponential decay of the wavefunctions, hence the homogeneous boundary condition in the finite computational domain for the Hartree potential may introduce large errors in the ground state energy \cite{davydov2016adaptive}. To efficiently reduce such errors, the multipole expansion of the Hartree potential \cite{jackson1999classical} can be used to generate an inhomogeneous boundary condition for the Poisson equation. It is noted that to keep using the homogeneous boundary condition, there also have been several works in the literature. The neutralization techniques have been studied \cite{lehtovaara2009all, pask2012linear} where an additional source term is introduced in the Poisson equation leading an $r^{-2}$ or even faster decay of the potential such that the homogeneous boundary condition makes sense. Another approach is adopting a sufficiently large computational domain for the Poisson equation as in \cite{motamarri2013higher}, wherein the diameter for the computational domain of the Poisson equation is around 100 times larger than that of the Kohn--Sham equation. In such a way, the zero boundary condition is adequate for the Hartree potential.
	
	As stated in \cite{schauer2013all, davydov2016adaptive}, one key issue in accurately approximating the Hartree potential is to choose an appropriate finite element space for the Poisson equation. Notably, the electron density, which represents the sum of squared wavefunctions and constitutes the right hand side of the Poisson equation, belongs to a finite element space twice the polynomial degree of that used for the wavefunctions. This suggests a preference for a larger finite element space in the quest to find the Hartree potential. Various approaches, including using a finer mesh with the same polynomial degree as the Kohn--Sham equation \cite{tsuchida1995electronic} or doubling the polynomial degree \cite{schauer2013all, davydov2016adaptive}, have shown improvements in the total energy calculations. However, it should be pointed out that these approaches are far from perfect, due to the reason that the approximate space for the Hartree potential in each approach is inherited from the one for wavefunctions. This space may not ideally suit the Hartree potential's distinct behaviors in comparison to wavefunctions. A potential remedy for creating a suitable space accommodating both the Hartree potential and the wavefunctions is the use of an \textit{a posteriori} error indicator \cite{verfurth2013posteriori}, considering contributions from both the Poisson and Kohn-Sham equations. While this approach can yield accurate results, it also introduces significant computational complexity, resulting from the substantial number of degrees of freedom defined on a single mesh, driven by differences not only in decay behavior but also in the magnitude of the error indicator between the two quantities. In light of these observations, a practical solution, that is, the \textit{multi-mesh adaptive method}, to simultaneously address accuracy and complexity concerns emerges through a combination of the aforementioned approaches. This involves the adaptive design of separate approximate spaces for wavefunctions and the Hartree potential. With such an idea, the numerical accuracy issue can be handled well since the approximate space for each variable is tailored according to its own feature, while the complexity is effectively reduced by replacing a large system of linear equations with two smaller ones.

	Distinct from the method that solves all the equations on the same mesh (hereafter we call such method \textit{single-mesh method}), the multi-mesh adaptive method (abbreviated as \textit{multi-mesh method}) is to solve different variables with different regularity on different meshes in the same computational domain. The quality of mesh associated with a certain variable can be improved from the mesh adaption process similar to the mesh adaption in the single-mesh method. In this study, the implementation of the multi-mesh method is based on the framework proposed in \cite{li2005multi}. Within this framework, mesh adaptation is achieved by locally refining or coarsening the element, a process known as $h$-adaptation. A similar idea can be found in \cite{dubcova2010space, solin2010monolithic} where $p$-adaption of the finite element space is also considered. The multi-mesh method has been successfully applied to various fields and models, such as in simulating the two- and three-dimensional dendritic growth \cite{hu2009multi, di2009computation}, the wetting and spreading problems \cite{di2009precursor}, the photonic band structure optimization \cite{wu2018multi}, the eigenvalue problems \cite{giani2021solving}, etc. With the multi-mesh method, chemical accuracy is able to be achieved since both wavefunctions and the Hartree potential are solved on different but qualified spaces.
	
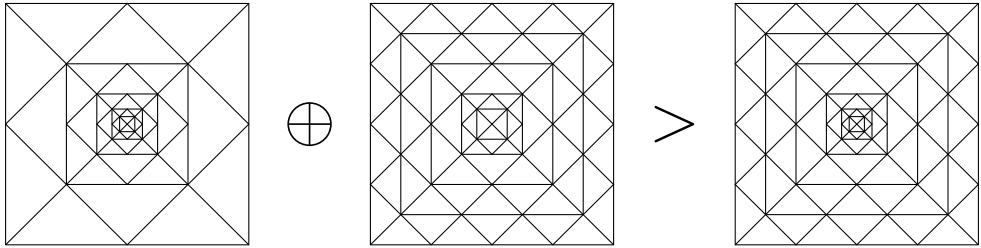
\begin{figure}[!h]
		\centering
		\begin{tikzpicture}[scale=0.2]
			
			\begin{scope}
				\draw (8,7)--(9,8) -- (8,9) -- (7,8)--cycle;
				\draw (0,0)--(16,0)--(16,16)--(0,16)--cycle;
				\draw (8,0)--(16,8)--(8,16)--(0,8)--cycle;
				\draw (4,4)--(12,4)--(12,12)--(4,12)--cycle;
				\draw (0,0)--(16,16);
				\draw (16,0)--(0,16);
				\draw (8,4)--(12,8)--(8,12)--(4,8)--cycle;
				\draw (6,6)--(10,6)--(10,10)--(6,10)--cycle;
				\draw (8,6)--(10,8) -- (8,10) -- (6,8)--cycle;
				\draw (7,7)--(9,7) -- (9,9) -- (7,9) --cycle;
				\draw (7.5,7.5)--(8.5,7.5) -- (8.5,8.5) -- (7.5,8.5) --cycle;

				\draw (20,8) node (plus) {{\Huge $\oplus$}};
				\draw[shift={(24,0)}] (0,0)--(16,0)--(16,16)--(0,16)--cycle;
				\draw[shift={(24,0)}] (8,0)--(16,8)--(8,16)--(0,8)--cycle;
				\draw[shift={(24,0)},fill=white] (4,4)--(12,4)--(12,12)--(4,12)--cycle;
				\draw[shift={(24,0)}] (0,0)--(16,16);
				\draw[shift={(24,0)}] (16,0)--(0,16);
				
				\draw[shift={(24,0)}] (2,2)--(14,2)--(14,14)--(2,14)--cycle;
				\draw[shift={(24,0)}] (16,12) -- (12,16)--(8,12)--(4,16)--(0,12)--(4,8)--(0,4)--(4,0)--(8,4)--(12,0)--(16,4)--(12,8)--(16,12);
				\draw[shift={(24,0)}]  (8,4)--(12,8)--(8,12)--(4,8)--cycle;
				\draw[shift={(24,0)}]  (6,6)--(10,6)--(10,10)--(6,10)--cycle;
				\draw[shift={(24,0)}]  (8,6)--(10,8) -- (8,10) -- (6,8)--cycle;
				\draw[shift={(24,0)}]  (7,7)--(9,7) -- (9,9) -- (7,9) --cycle;
				
				\draw (44,8) node (plus) {{\Huge $>$}};
				\draw[shift={(48,0)}] (0,0)--(16,0)--(16,16)--(0,16)--cycle;
				\draw[shift={(48,0)}] (8,0)--(16,8)--(8,16)--(0,8)--cycle;
				\draw[shift={(48,0)}] (4,4)--(12,4)--(12,12)--(4,12)--cycle;
				\draw[shift={(48,0)}] (0,0)--(16,16);
				\draw[shift={(48,0)}] (16,0)--(0,16);
				
				\draw[shift={(48,0)}] (2,2)--(14,2)--(14,14)--(2,14)--cycle;
				\draw[shift={(48,0)}] (16,12) -- (12,16)--(8,12)--(4,16)--(0,12)--(4,8)--(0,4)--(4,0)--(8,4)--(12,0)--(16,4)--(12,8)--(16,12);
				\draw[shift={(48,0)}] (8,4)--(12,8)--(8,12)--(4,8)--cycle;
				\draw[shift={(48,0)}] (6,6)--(10,6)--(10,10)--(6,10)--cycle;
				
				\draw[shift={(48,0)}] (8,7)--(9,8) -- (8,9) -- (7,8)--cycle;
				\draw[shift={(48,0)}] (8,6)--(10,8) -- (8,10) -- (6,8)--cycle;
				\draw[shift={(48,0)}] (7,7)--(9,7) -- (9,9) -- (7,9) --cycle;
				\draw[shift={(48,0)}] (7.5,7.5)--(8.5,7.5) -- (8.5,8.5) -- (7.5,8.5) --cycle;
				
			\end{scope}
		\end{tikzpicture}
		\caption{A sample for multi-mesh method. The left two images show the meshes used in the multi-mesh method. The right one shows the mesh used in the single-mesh method. The number of the mesh grids for the three meshes are 37, 49, and 57, respectively.   \label{fig:mul-sample}}
	\end{figure}

	An anticipated advantage of the multi-mesh method, when compared to the single-mesh method, is the potential to reduce computational costs in achieving chemical accuracy. This can be explained by a sample as displayed in \Cref{fig:mul-sample}. The first mesh can be viewed as the mesh for solving the Kohn--Sham equation, with a dense grid distribution near the singularity and a sparse distribution for regions far from it. The middle mesh is used to solve the Hartree potential, and due to its smooth behavior and low decay speed, the mesh grid distribution remains uniform. Finally, the last mesh, formed without employing the multi-mesh strategy, should be able to capture both the singularity and the low decay of the Hartree potential. The number of mesh grids for these three meshes is 37, 49, and 57, respectively. Assume that the linear finite element method is adopted. In the multi-mesh method, a generalized eigenvalue problem $Ax=\lambda Bx$ with $A,B \in \mathbb{R}^ {37\times 37}$  and a linear system $Sx=b$ with $S\in \mathbb{R}^{49\times 49}$. While for the finite element method using one mesh, the eigenvalue problem and the linear system are of the same size $A,B,S\in \mathbb{R}^{57\times 57}$. Apparently, the computational cost for solving these two problems in the multi-mesh method is much less than the single-mesh method in this sample. 
	
	Nevertheless, the application of the multi-mesh method is accompanied by several challenges. The first of these challenges is the effective management of mesh grids, requiring the capacity to perform local grid refinements or coarsening in response to specific problem demands. This management process must seamlessly facilitate the update of solutions from the previous mesh to the new one. The second challenge centers on the critical necessity for efficient and accurate multi-mesh communication, particularly when computing integrals. The establishment of smooth communication is indispensable for maintaining consistency and accuracy across different meshes. To address these challenges, we introduce the hierarchical geometry tree data structure. This tree-based approach facilitates the natural and efficient manner in the management of the mesh grids. Furthermore, in order to prevent any loss of accuracy during communications across various meshes, we adopt a strategy that maximizes the utilization of quadrature points in numerical integrals, relying solely on interpolation without the need for a projection process. This approach ensures the preservation of numerical accuracy throughout the communication process.
	
	By employing the multi-mesh approach, we establish a multi-mesh adaptive finite element framework for the all-electron density functional theory. With the additional treat on the Hartree potential, systematical convergence of the total energy to chemical accuracy is observed and the computational cost is less compared with the single-mesh method given the same accuracy. The mesh adaption is based on the residual-based \textit{a posteriori} error estimation of the Kohn--Sham equation and the Poisson equation whose solution is the Hartree potential. The ability to achieve chemical accuracy is verified by a series of numerical examples of atoms and molecules. By a detailed comparison of computational time between the single-mesh method and the multi-mesh method, the efficiency of the multi-mesh adaptive method is justified.

	The organization of this paper is listed below. In Section 2, we introduce the Kohn--Sham equation, the Poisson equation, and the finite element discretizations of these two equations. Then in Section 3, the single-mesh adaptive finite element framework for Kohn--Sham density function theory is reviewed and tested on the Helium atom example. Followed by Section 4 in which the multi-mesh adaptive method is demonstrated and examined also on the Helium example. In Section 5, the efficiency and accuracy of the presented method are discussed in detail from numerical examples. Finally, this paper ends with the conclusion in Section 6.
	
	\section{The Kohn--Sham equation and finite element discretization}
	\subsection{Kohn--Sham density functional theory}
	\label{sec:ksdft}
	The Kohn--Sham equation for a $p$-electron system is read as the following eigenvalue problem
	\begin{equation}\label{eq:KS}
		\left\{ 
		\begin{array}{lr}
			\hat{H}\psi_l(\mathbf{r}) = \varepsilon_l\psi_l(\mathbf{r}),& l=1,2,\dots ,p,\\
			\displaystyle\int_{\mathbb{R}^3} \psi_l\psi_{l'} \,\mathrm{d}\mathbf{r} = \delta_{ll'},
			&l,l'=1,2,\dots ,p, 
		\end{array} \right. 
	\end{equation}
	where $(\varepsilon_l,\psi_l)$ is the $l$-th eigenpair, $\delta_{ll'}$ is the Kronecker delta, and $\hat{H}$ stands for the Hamiltonian operator. We denote $\rho(\mathbf{r}) = \sum_{l=1}^p|\psi_l|^2(\mathbf{r})$ as the electron density. Specifically, $\hat{H}$ consists of the following four terms
	\begin{equation}\label{eq:Hamiltonian}
		\hat{H}([\rho];\mathbf{r}) = -\frac{1}{2}\nabla_{\mathbf{r}}^2 + V_{\mathrm{ext}}(\mathbf{r}) + V_{\mathrm{Har}}([\rho];\mathbf{r}) + V_{\mathrm{xc}}([\rho];\mathbf{r}),
	\end{equation}
	where the notation $V([\rho];\mathbf{r})$ implies that $V$ is a functional of the electron density
	$ \rho$. The first term  $-\nabla^2/2$ in $\hat{H}$ is the kinetic operator. The second term in $\hat{H}$ describes the Coulomb external potential due to the nuclei which takes the form
	\begin{equation*}
		V_{\mathrm{ext}}(\mathbf{r}) = -\sum_{j=1}^{M} \frac{Z_j}{\lvert
			\mathbf{r}-\mathbf{R}_j\rvert},
	\end{equation*}
	where $M$ is the number of nuclei. The third term is the Hartree potential describing the Coulomb repulsion among the electrons
	\begin{equation}\label{eq:har}
		V_{\mathrm{Har}}([\rho];\mathbf{r})=\int_{\mathbb{R}^{3}} \frac{\rho(\mathbf{r'})}
		{\lvert \mathbf{r}-\mathbf{r'} \lvert} \,\mathrm{d}\mathbf{r'}.
	\end{equation}
	The last term $V_{\mathrm{xc}}$ stands for the exchange-correlation potential, which is caused by the Pauli exclusion principle and other non-classical Coulomb interactions. Note that the analytical expression for the exchange-correlation term is unknown and therefore an approximation is needed. Specifically, the local density approximation (LDA) from the library Libxc \cite{marques2012libxc} with the slater exchange potential and the Vosko-Wilk-Nusair (VWN4) \cite{vosko1980accurate} is adopted in this work.  
	
	Note that direct evaluation of the Hartree potential \eqref{eq:har} requires computational cost $\mathcal{O}(N^2)$ with $N$ being the number of grid points on the computational domain $\Omega$. For simplicity, we denote $\phi = V_{\mathrm{Har}}(\mathbf{r})$ hereafter. In this paper, the Hartree potential is obtained by solving the Poisson equation
	\begin{equation}\label{eq:poisson}
		\left\{
		\begin{aligned}
			-\nabla^2\phi(\mathbf{r}) = 4\pi \rho(\mathbf{r}),&~ \text{in } \Omega,\\
			\phi(\mathbf{r}) = \phi_{\partial \Omega}(\mathbf{r}),&~ \text{on } \partial \Omega.
		\end{aligned}
		\right.
	\end{equation}
	The boundary value $\phi_{\partial \Omega}(\mathbf{r})$ is evaluated by the multipole expansion method. Specifically, the following approximation is used
	\begin{equation*}
		\begin{aligned}
			\left.
			\phi(\mathbf{r})\right|_{\mathbf{r} \in \partial \Omega} \approx & \frac{1}{\left|\mathbf{r}-\mathbf{r}^{\prime \prime}\right|} \int_{\Omega} \rho\left(\mathbf{r}^{\prime}\right) \,\mathrm{d} \mathbf{r}^{\prime}+\sum_{i=1,2,3} p_{i} \cdot \frac{r^{i}-r^{\prime \prime, i}}{\left|\mathbf{r}-\mathbf{r}^{\prime \prime}\right|^{3}} +\sum_{i, j=1,2,3} q_{i j} \cdot \frac{3\left(r^{i}-r^{\prime \prime, i}\right)\left(r^{j}-r^{\prime \prime, j}\right)-\delta_{i j}\left|\mathbf{r}-\mathbf{r}^{\prime \prime}\right|^{2}}{\left|\mathbf{r}-\mathbf{r}^{\prime \prime}\right|^{5}},
		\end{aligned}
	\end{equation*}
	where
	\begin{equation*}
		p_{i}=\int_{\Omega} \rho\left(\mathbf{r}^{\prime}\right)\left(r^{\prime, i}-r^{\prime \prime, i}\right) \,\mathrm{d} \mathbf{r}^{\prime}, \quad q_{i j}=\int_{\Omega} \frac{1}{2} \rho\left(\mathbf{r}^{\prime}\right)\left(r^{\prime, i}-r^{\prime \prime, i}\right)\left(r^{\prime, j}-r^{\prime \prime, j}\right) \,\mathrm{d} \mathbf{r}^{\prime}.
	\end{equation*}
	In the above expressions, $\mathbf{r}^{\prime\prime}$ stands for an arbitrary point in $\Omega$. In the simulations, we choose it to be
	\begin{equation*}
		\mathbf{r}^{\prime\prime} = \frac{\int \mathbf{r} \rho(\mathbf{r}) \,d\mathbf{r}}{\int  \rho(\mathbf{r}) \,d\mathbf{r}}.
	\end{equation*}
	
	\subsection{Finite element discretization}
	In practical simulations, a bounded polyhedral domain $\Omega \subset \mathbb{R}^3$ is served as the computational domain. Assume that the finite element space $V_h$ is constructed on $\Omega$ and the finite element basis of $V_h$ is denoted as $\{\varphi_1,\dots,\varphi_n\}$ with $n$ being the dimension of the space. Then the wavefunction $\psi_i$ can be approximated as $\psi_i^h$ on $V_h$ via
	\begin{equation}\label{eq:x-fem}
		\psi_{i}^{h}(\mathbf{r})=\sum_{k=1}^{n} X_{k, i} \varphi_{k}, \quad X_{k, i}=\psi_{i}\left(\mathbf{r}_{k}\right), \quad X \in \mathbb{R}^{n \times p},
	\end{equation}
	where $\mathbf{r}_k$ denotes the node corresponding to the $i$-th basis function. As a result, to find the approximation of the wavefunctions $\{\psi_i\}$ on $V_h$ is to find $X$, i.e., the value of $\psi_i$ at each node $\mathbf{r}_k$.
	
	In the finite element space $V_h$, the variation form for the KS equation \eqref{eq:KS} becomes: find $(\psi_i^h,\varepsilon_i^h)_{i=1,\dots,p} \in V_{h}\times \mathbb{R}$, such that
	\begin{equation*}
		\int_{\Omega}\varphi\hat{H}\psi_i^h
		\,\mathrm{d} \mathbf{r} = \varepsilon_i^h\int_{\Omega} \varphi \psi_i^h \,\mathrm{d} \mathbf{r},\quad \forall \varphi \in V_h.
	\end{equation*}
	By letting $\varphi$ be the finite element basis function and inserting \eqref{eq:x-fem} to the above variational form, we have the following discrete eigenvalue problem
	\begin{equation}\label{eq:evp-vh}
		A X = \varepsilon M Y.
	\end{equation}
	Here $A$ and $M$ are symmetric matrices with the entries
	\begin{align}
		\label{eq:matA}
		A_{i,j} &=  \int_{\Omega} \frac{1}{2}\nabla \varphi_j\cdot \nabla \varphi_i + \big( V_{\mathrm{ext}}+\phi+V_{\mathrm{xc}} \big)\varphi_j\varphi_i \,\mathrm{d} \mathbf{r},\\
		\label{eq:matM}
		M_{i,j} &= \int_{\Omega} \varphi_j\varphi_i \,\mathrm{d} \mathbf{r}.
	\end{align}
	
	Similarly, we can obtain the linear system for the Poisson equation \eqref{eq:poisson} on $V_h$:
	\begin{equation}\label{eq:ls-har}
		S \Phi = \mathbf{b},
	\end{equation}
	where $S$ is the stiff matrix with the entry
	\begin{equation}\label{eq:matS}
		S_{i,j} =  \int_{\Omega} \frac{1}{2}\nabla \varphi_j\cdot \nabla \varphi_i \,\mathrm{d} \mathbf{r},
	\end{equation}
	and the right hand side $\mathbf{b}$ and the discretized $\phi^h$ 
	\begin{equation*}
		\mathbf{b}_i =  \int_{\Omega} 4\pi \rho(\mathbf{r}) \varphi_i \,\mathrm{d} \mathbf{r},\quad
		\phi^h(\mathbf{r}) = \sum_{k=1}^n \Phi_k \varphi_k,\quad \Phi \in \mathbb{R}^n.
	\end{equation*}
	
	Owing to the singularity in the Hamiltonian \eqref{eq:Hamiltonian}, a uniform finite element discretization would lead to a large number of mesh grids to achieve chemical accuracy. Hence the adaptive mesh method is necessary to efficiently solve the Kohn--Sham equation, which will be discussed in the next section.
	
	\section{The adaptive finite element method}
	In this section, we will begin by introducing the adaptive finite element method that relies on the \emph{a posteriori} error estimates \cite{verfurth2013posteriori}. Subsequently, we will present and compare two error indicators used to facilitate mesh adaptation for solving the Kohn--Sham equation \eqref{eq:KS}. The sole distinction between these two error estimators is their utilization of the error involved in solving the Hartree potential.
	
	\subsection{The adaptive method based on the residual \emph{a posteriori} error estimation}
	The adaptive mesh techniques offer enhanced numerical accuracy while requiring fewer mesh grids compared to uniform mesh methods. In this study, our primary focus lies on the $h$-adaptive methods, which involve local refinement and/or coarsening of mesh grids. An important aspect of $h$-adaptive methods is the determination of an error indicator. Generally speaking, error indicators identify regions within the domain that necessitate local refinement or coarsening, and they are typically derived from \emph{a posteriori} error estimations \cite{verfurth2013posteriori}. When there is only one orbital in the system, it is natural to generate the indicator based on information from that specific orbital. However, when there are multiple orbitals in the system, generating the indicator solely from an individual orbital is no longer advisable. This is because every orbital in the system is expected to be well-resolved using the mesh grids after mesh adaptation. To address this, we adopt the strategy proposed in \cite{bao2012h} for indicator generation. First, indicators are individually generated for each orbital using a specific method. Then, normalization is applied to each indicator. The final indicator is obtained by combining these normalized indicators.
	
	Specifically, based on the \emph{a posteriori} error estimates \cite{verfurth2013posteriori},  the residual-based \emph{a posterirori} error indicator  for the KS equation \eqref{eq:KS} in the element $\mathcal{T}_K$ can be defined as
	\begin{equation}\label{eq:indi-ks}
		\eta_{K,\mathrm{KS}} =\left(h_{K}^{2}\sum_{l=1}^p\big\|\mathbb{R}_{K,\mathrm{KS}}(\psi_l)\big\|_{K}^{2}+\sum_{e \subset \partial \mathcal{T}_{K}} \frac{1}{2} h_{e}\sum_{l=1}^p\big\|\mathbb{J}_{e}(\psi_l)\big\|_{e}^{2}\right)^{\frac{1}{2}},
	\end{equation}
	where $h_K$ represents the largest length of the edges of the element $\mathcal{T}_K$, $h_{e}$ stands for the largest length of the common face $e$ of $\mathcal{T}_K$ and $\mathcal{T}_J$, $\mathbb{R}_{K,\mathrm{KS}}(\psi)$ and $\mathbb{J}_{e}(\psi)$ are the residual and jump term on the element $\mathcal{T}_K$, whose formulations are written as
	\begin{equation*}
		\left\{
		\begin{aligned}
			\mathbb{R}_{K,\mathrm{KS}}(\psi_l)&=
			\hat{H} \psi_l- \varepsilon_lM\psi_l,\\
			\mathbb{J}_{e}(\psi_l) &= (\nabla
			\psi_l\bigm|_{\mathcal{T}_{K}} - \nabla\psi_l
			\bigm|_{\mathcal{T}_{J}}) \cdot \mathbf{n}_{e},
		\end{aligned}
		\right.
	\end{equation*}
	where $\mathbf{n}_{e}$ is the out normal vector on the face $e$ w.r.t the element $\mathcal{T}_K$. The definition of the norms are
	\begin{equation*}
		\left\| f(x) \right\|_K = \left(\int_K (f(x))^2dx\right)^{\frac{1}{2}},~~
		\left\| f(x) \right\|_{e} = \left(\int_{e} (f(x))^2dx\right)^{\frac{1}{2}}.
	\end{equation*}
	The indicator \eqref{eq:indi-ks} involves the error arising from the Kohn--Sham equation \eqref{eq:KS} and is usually directly adopted to guide the mesh adaption. 
	
	An aspect that the indicator \eqref{eq:indi-ks} may overlook is the error associated with the Hartree potential. It is important to emphasize the accurate approximation of the Hartree potential, as it is a crucial component of the Hamiltonian \eqref{eq:Hamiltonian} and contributes to the Hartree potential energy. As the Hartree potential is obtained by solving the Poisson equation \eqref{eq:poisson}, a similar approach can be employed to generate an error indicator specifically for the Hartree potential, just as done for the Kohn--Sham equation. Specifically, the error indicator for the Hartree potential can be defined as
	\begin{equation}\label{eq:indi-poisson}
		\eta_{K,\mathrm{Har}} =\left(h_{K}^{2}\big\|\mathbb{R}_{K,\mathrm{Har}}(\phi)\big\|_{K}^{2}+\sum_{e \subset \partial \mathcal{T}_{K}} \frac{1}{2} h_{e}\big\|\mathbb{J}_{e}(\phi)\big\|_{e}^{2}\right)^{\frac{1}{2}},
	\end{equation}
	where the residual part is defined as $\mathbb{R}_{K,\mathrm{Har}}(\phi)= \nabla^2\phi + 4\pi\rho$. Based on the  analysis above,  the second indicator for solving the Kohn--Sham equation can be designed as 
	\begin{equation} \label{eq:indi-kshar}
		\eta_{K,\mathrm{KS+Har}} = \sqrt{\eta_{K,\mathrm{Har}}^2 + \eta_{K,\mathrm{KS}}^2}.
	\end{equation}
	The aim of using this error indicator is to generate a mesh on which both the wavefunctions and the Hartree potential are approximated well. 
	
	The mesh adaptation process, along with the solution of the Kohn--Sham equation, can be carried out using either the first \eqref{eq:indi-ks} or the second error indicator \eqref{eq:indi-kshar}. It is important to note that the adaptive algorithms utilizing these two different indicators are essentially identical, except for the choice of the indicator. Moreover, all simulations are performed on a single mesh simultaneously, i.e., the Poisson equation \eqref{eq:poisson} for the Hartree potential is discretized and solved on the same finite element space for the Kohn--Sham equation. Consequently, we present the adaptive finite element method for solving the Kohn--Sham equation, outlined in Algorithm \ref{alg:single}, and refer to this approach as the \emph{single-mesh adaptive method}. Briefly, the adaptive algorithm consists of an outer iteration and an inner iteration. In the inner iteration, it is an SCF method for solving the nonlinear Kohn--Sham equation. Meanwhile, it is the mesh adaption procedure in the outer iteration. In the next subsection, we would like to show an example using the single mesh adaptive method with the first indicator \eqref{eq:indi-ks}. 
	\begin{algorithm}
		\caption{Single-mesh adaptive method. \label{alg:single}}
		\KwIn{Initial mesh $\mathcal{T}^{(0)}$, initial electron density $\rho$, initial energy $E= 0$, energy tolerance $tol_1$, density tolerance $tol_2$.}
		
		\While{$|E-E_{old}|<tol_1$}{
			$E_{old} = E$;
			
			\While{$\Vert\rho-\rho_{old}\Vert<tol_2$}{
				
				$\rho_{old} = \rho$;
				
				Calculate the Hartree potential $\phi$;
				
				Generate the Hamiltonian matrix $A$;
				
				Solve the the eigenvalue problem $AX = \varepsilon MX$;
				
				Update the electron density $\rho$;
			}
			
			Calculate the ground state energy $E$;
			
			Mesh adaption based on the error indicator.
		}
		
		\KwOut{Total ground state energy $E$}
		
	\end{algorithm}
	\subsection{An issue for numerical solutions towards chemical accuracy}
	\subsubsection{Single-mesh adaptive method using indicator \eqref{eq:indi-ks}}
	To assess the convergence and behavior of the single mesh adaptive finite element method with the error indicator \eqref{eq:indi-ks}, we conduct a series of eight experiments for the helium atom, each involving a different number of mesh grids, controlled by varying adaption tolerance values. In this example, the referenced value is obtained from the state-of-art software \texttt{NWChem} using the \emph{aug-cc-pv6z} basis set \cite{valiev2010nwchem}. The hardware and software configurations of the experiment are introduced in the beginning of \Cref{sec:numEx}.
	
\begin{figure}[!htp]
		\centering
		\includegraphics[width=0.45\linewidth]{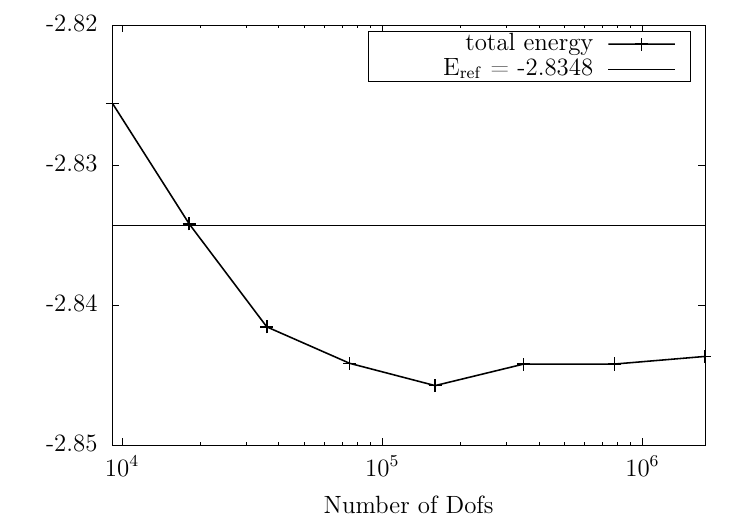}~
		\includegraphics[width=0.45\linewidth]{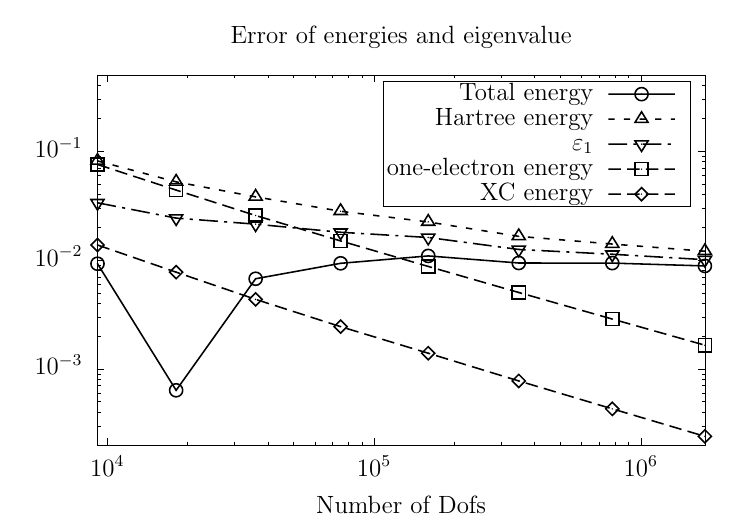}
		\caption{Convergence history of the total enegy (left), absolute errors of energies and eigenvalue (right) for the adaptive method using error indicator \eqref{eq:indi-ks}. \label{fig:he-aks}}
	\end{figure}

	The results are displayed in the  \Cref{fig:he-aks}. As the tolerance for error indicator decreases, the number of degrees of freedom increases. The convergence of the total energy would serve as evidence of the superiority of the adaptive method compared to the method employing a uniform mesh partition. However, it is evident that the total energy does not converge to the reference result observed from the left of \Cref{fig:he-aks}. Even in the mesh generated from the smallest tolerance, which consists of 1,737,569 mesh grids, a discrepancy of 0.01 Hartree in the energy is still observed, indicating that the adaptive method does not achieve the desired convergence.

	As discussed earlier, the error in the total energy primarily stems from the inaccurate approximation of the Hartree energy. This is confirmed by comparing the Hartree energy with the reference value $E_\mathrm{Har,ref}=1.9961$ Hartree, as depicted on the right side of  \Cref{fig:he-aks}. 
	It is observed that although the error in the Hartree energy diminishes with respect to the number of Dofs, on the finest mesh, the Hartree energy is found to be 0.01 Hartree lower than the reference value, resulting in a 0.01 Hartree error in the total energy. Furthermore, an imprecise approximation of the Hartree potential can also introduce a 0.01 Hartree error in the eigenvalue, as depicted on the right side of \Cref{fig:he-aks}. As a comparison, it can be observed that the errors in the one-electron energy and the exchange-correlation energy systematically decrease as the number of Dofs increases.

\begin{figure}[!h]
		\centering
		\includegraphics[width=0.24\linewidth]{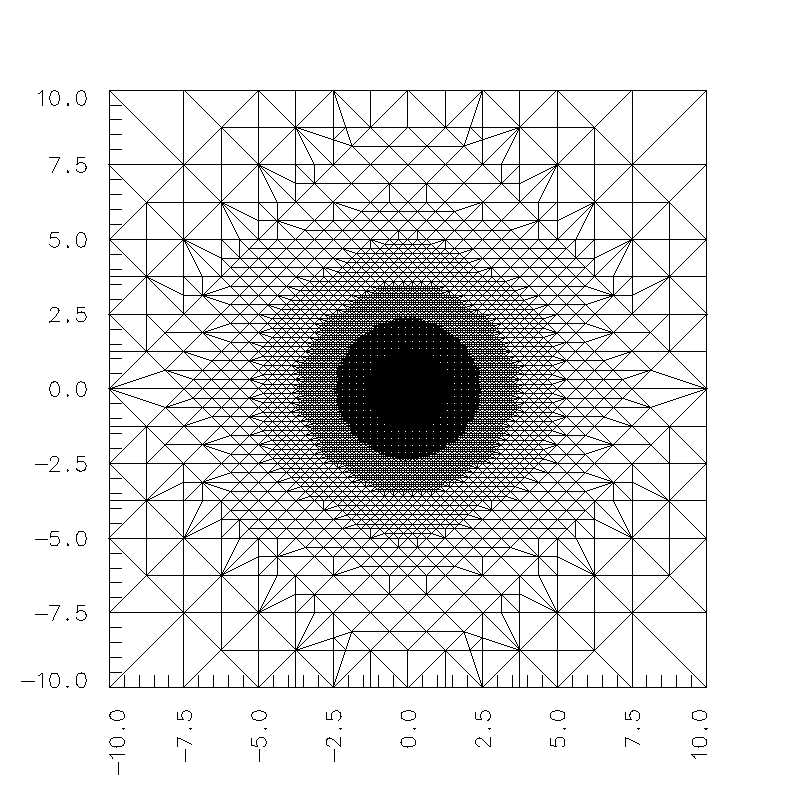}
		\includegraphics[width=0.24\linewidth]{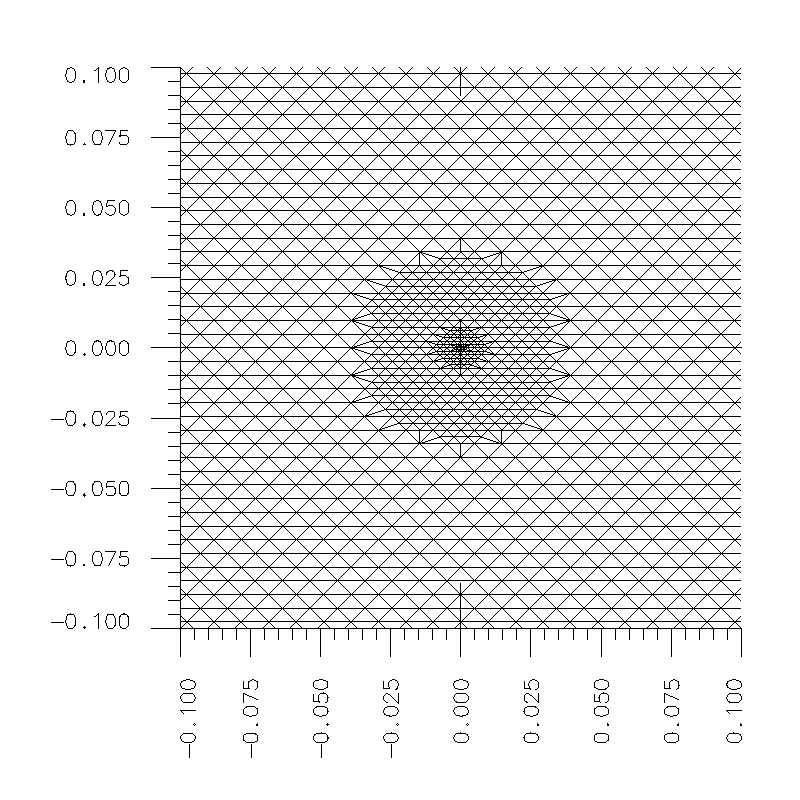}
		\includegraphics[width=0.24\linewidth]{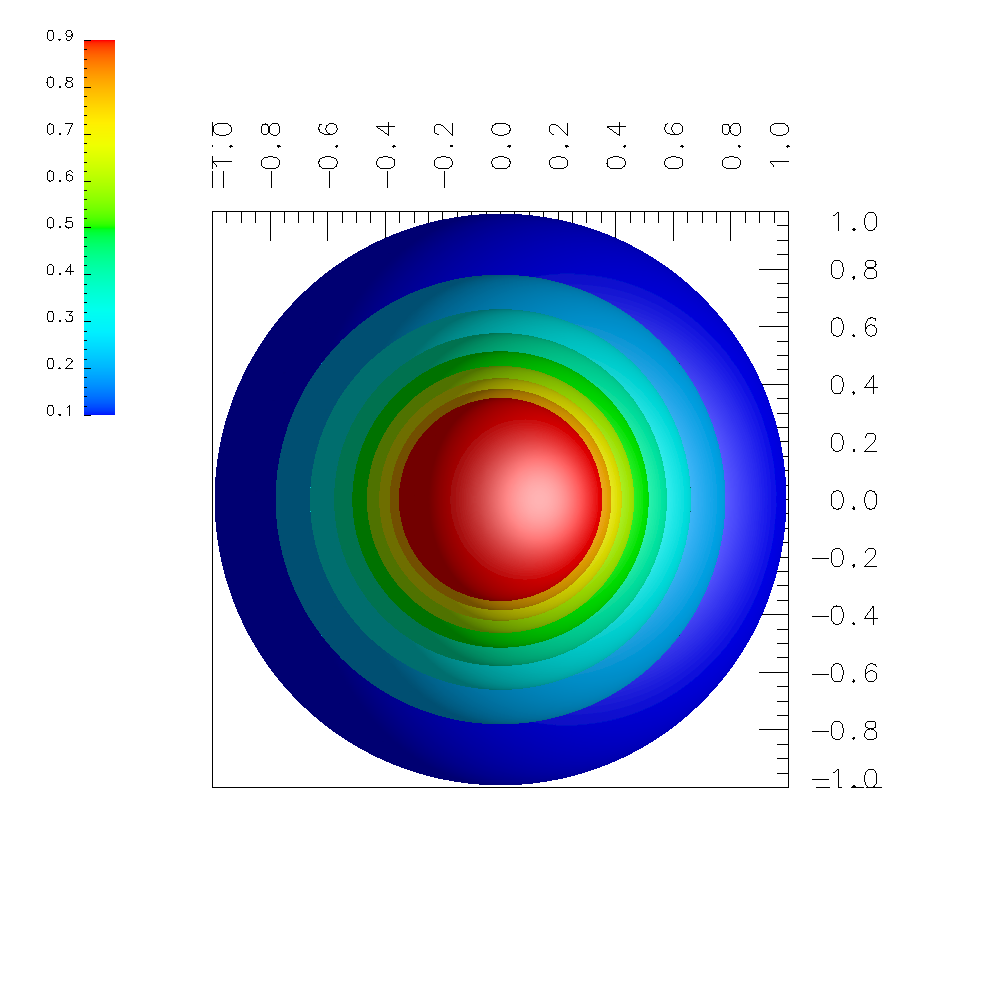}
		\includegraphics[width=0.24\linewidth]{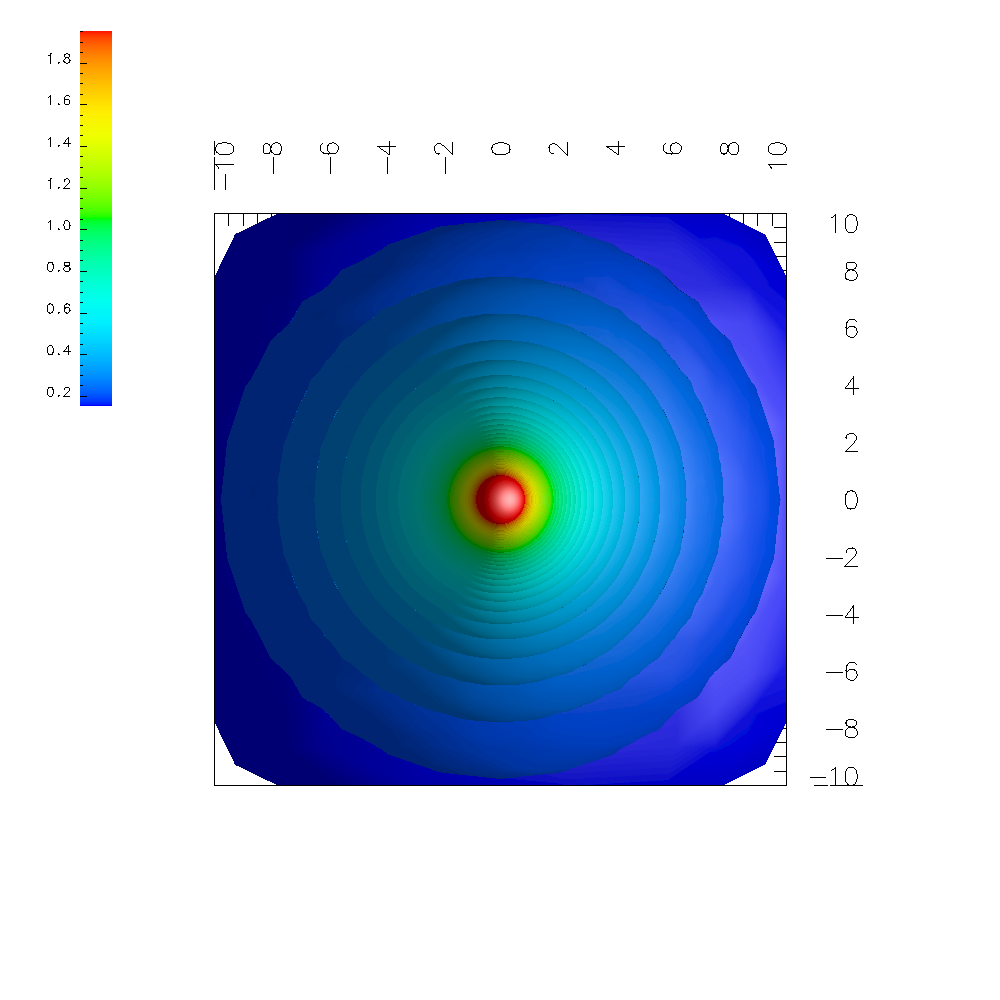}\\
		\includegraphics[width=0.24\linewidth]{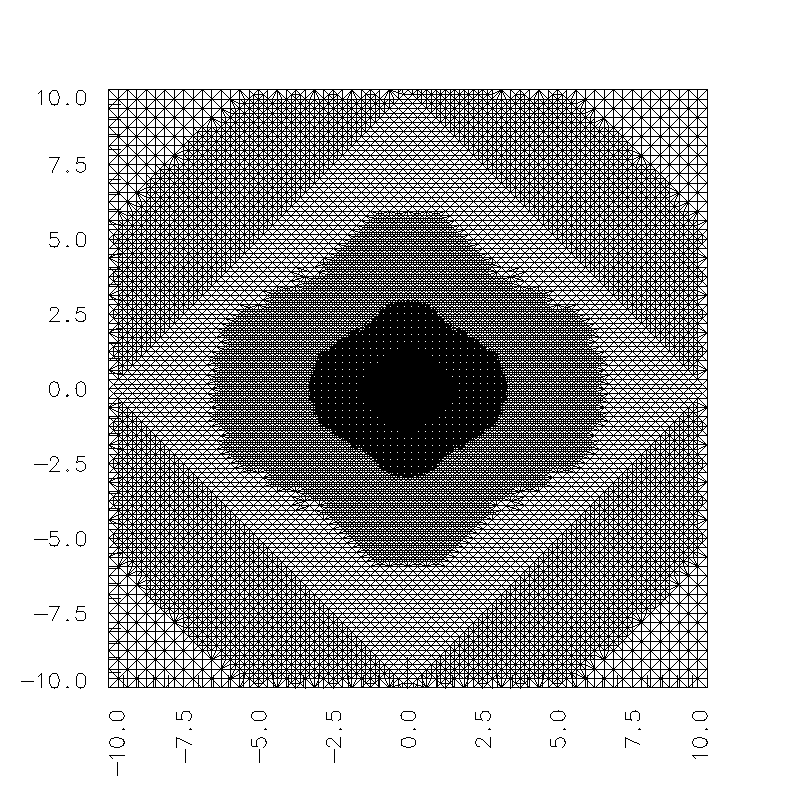}
		\includegraphics[width=0.24\linewidth]{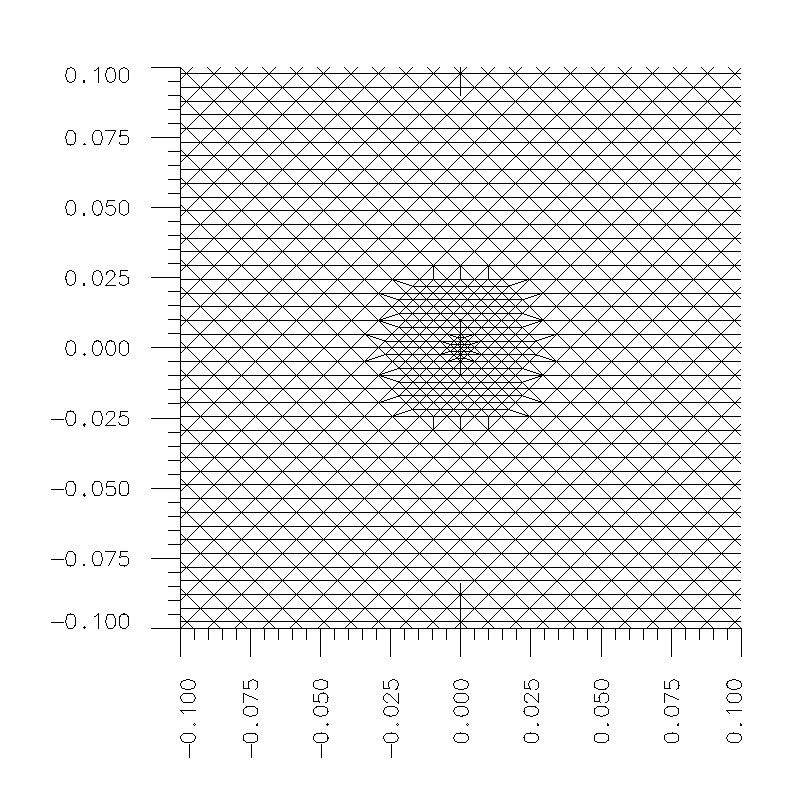}
		\includegraphics[width=0.24\linewidth]{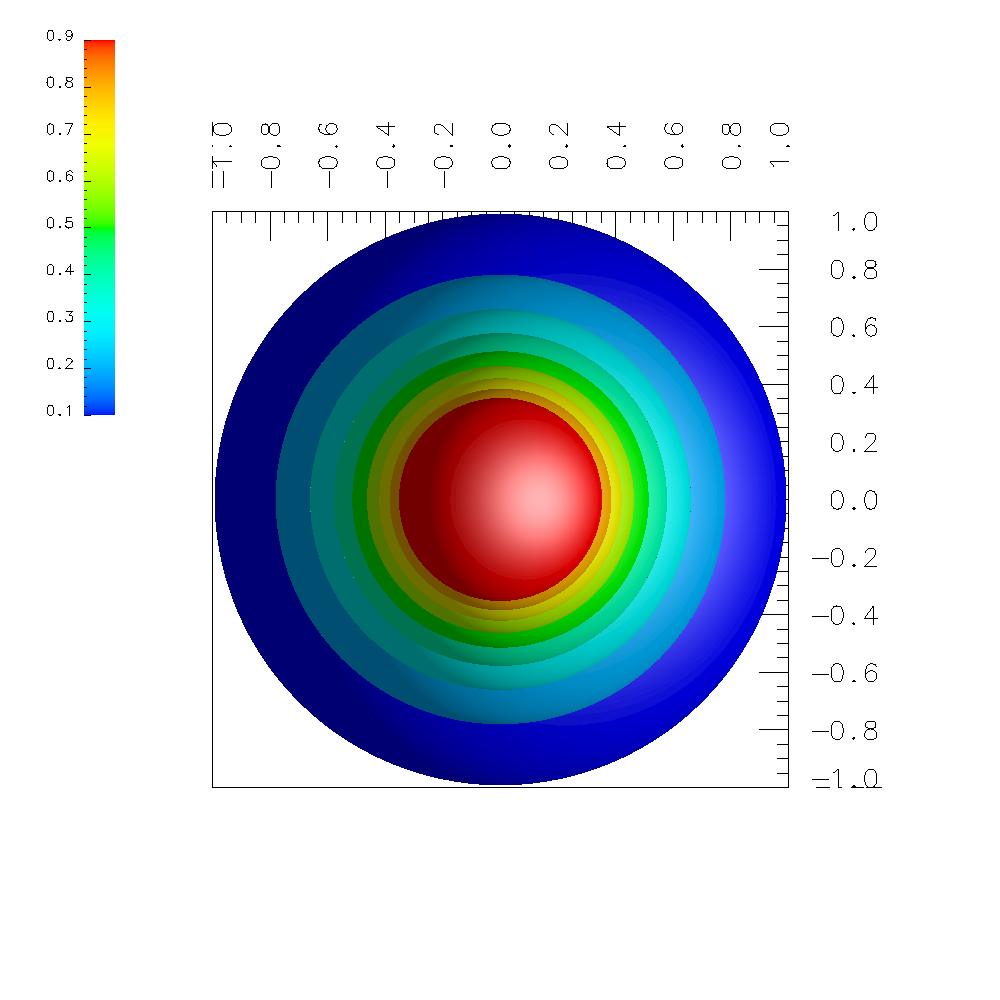}
		\includegraphics[width=0.24\linewidth]{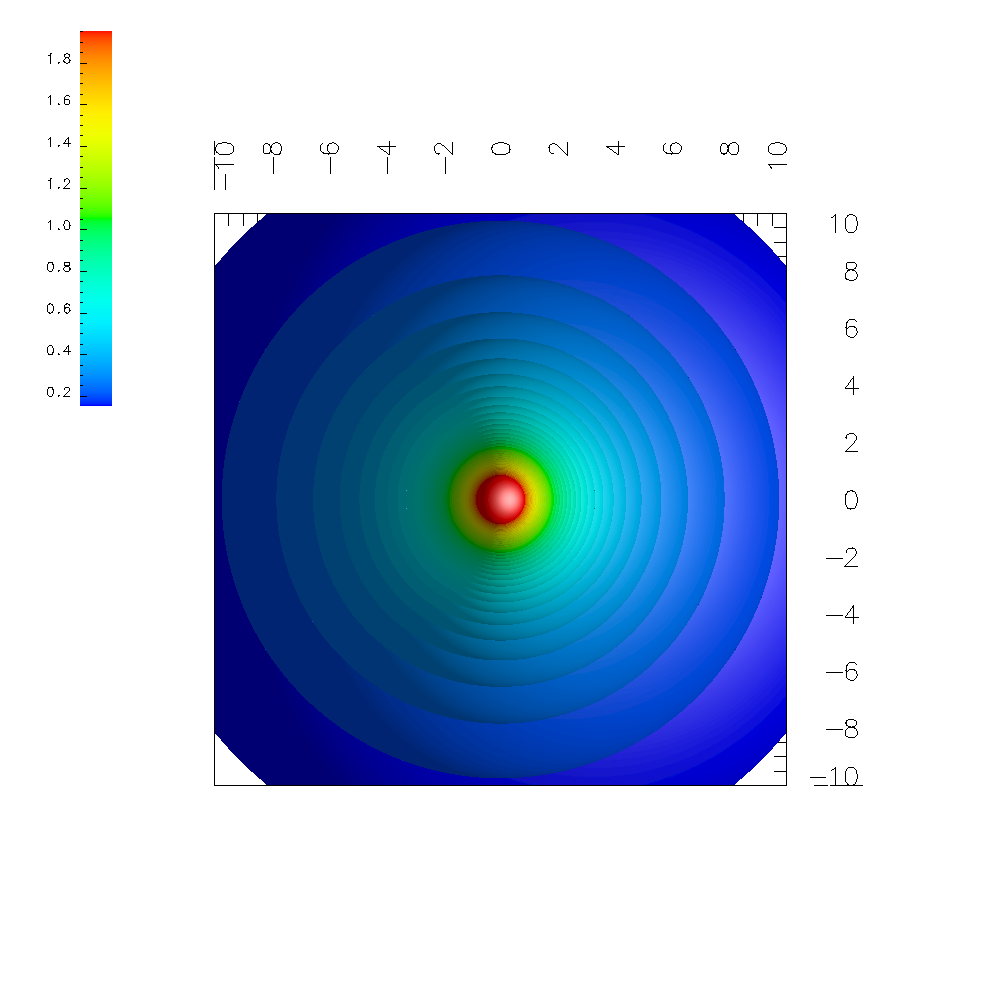}
		\caption{Top row: results from a single mesh adaptive method using the indicator \eqref{eq:indi-ks}, with total 1,765,990 mesh grids, i.e., Global mesh and zoomed-in mesh on the sliced $X$-$Y$ plane (left two), density profile (third one), and Hartree potential profile (fourth one). Bottom row:  corresponding results using the indicator \eqref{eq:indi-kshar}, with total 2,847,807 mesh grids.\label{fig:he-aks-mesh}}
	\end{figure}

	The reason behind the inaccurate approximation of the Hartree potential can be elucidated by referring to  \Cref{fig:he-aks-mesh} (upper row), which displays the results obtained from the finest mesh consisting of 1,765,990 mesh grids. In the left two columns, the global and local mesh distributions are depicted. It is evident that the mesh grid density is considerably high around the singularity located at the origin, with the smallest mesh size being approximately 0.001 Bohr. Conversely, in regions distant from the singularity, the mesh grid distribution is sparse. Specifically, near the boundary, the largest mesh size can exceed 2 Bohr. Such a mesh distribution proves to be suitable for representing wavefunctions and electron density, as illustrated in the third column of \Cref{fig:he-aks-mesh} (upper row), since both quantities exhibit exponential decay, and the region where the density exceeds 0.1 is confined to a small interval of $[-1, 1]^3$ for the helium atom. However, in the fourth column of Figure \ref{fig:he-aks-mesh} (upper row), it is apparent that the Hartree potential exhibits a much slower decay behavior, with potential values surpassing 0.1 throughout the entire computational domain of $[-10, 10]^3$. Furthermore, the contours near the boundary oscillate and are not smooth. As a result, a mesh size as large as 2 Bohr is evidently insufficient for capturing the variations in the Hartree potential accurately. In order to enhance the numerical accuracy of the Kohn--Sham equation, it is essential to obtain a more precise approximation of the Hartree potential.

	\subsubsection{Single mesh adaptive method using indicator \eqref{eq:indi-kshar}}
	To achieve a more accurate Hartree potential, a direct approach is to incorporate the corresponding error indicator \eqref{eq:indi-poisson} with the indicator of the Kohn--Sham equation \eqref{eq:indi-ks}. This combination results in the second indicator \eqref{eq:indi-kshar}. Consequently, we apply the single mesh adaptive method again to the Helium example, but this time utilizing the second indicator \eqref{eq:indi-kshar}.
	
\begin{figure}[h]
		\centering
		\includegraphics[width=0.45\linewidth]{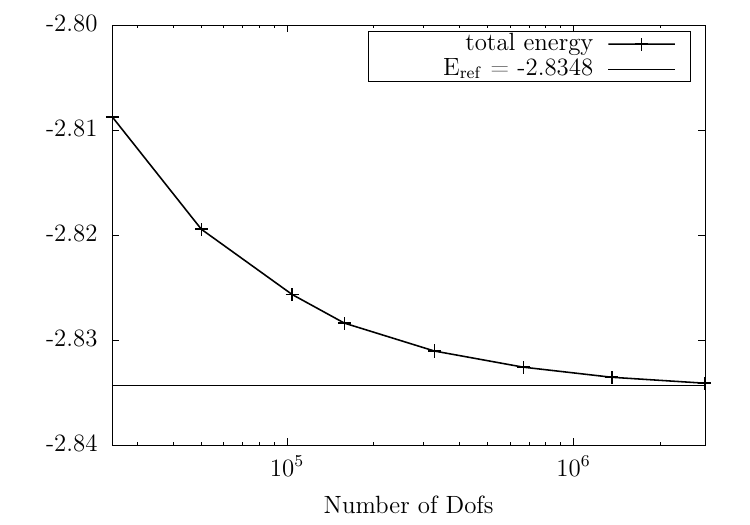}~
		\includegraphics[width=0.45\linewidth]{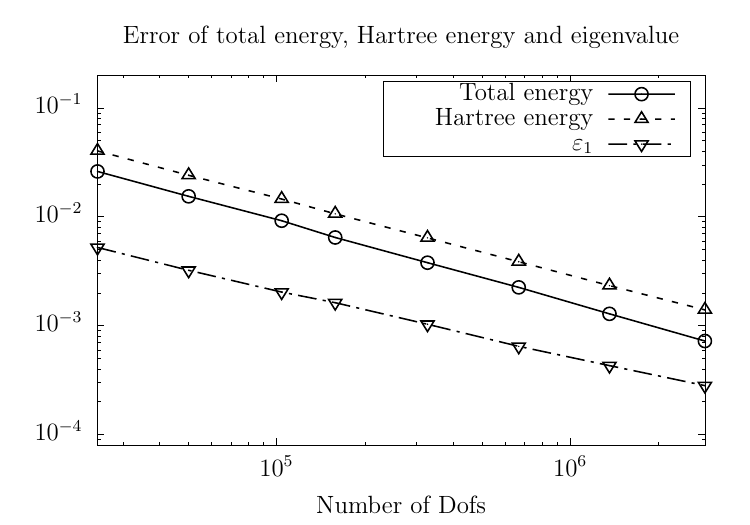}
		\caption{Convergence history of the total energy (left), absolute errors of energies and eigenvalue (right) for the adaptive method using error indicator \eqref{eq:indi-kshar}.\label{fig:he-akshar}}
	\end{figure}

	The convergence of energies and eigenvalues is depicted in \Cref{fig:he-akshar}. Notably, there is a systematic convergence observed in both the total energy and the Hartree potential energy. Moreover, the eigenvalue demonstrates convergence towards the reference value. On the finest mesh, which consists of 2,874,807 grid points, chemical accuracy is achieved. Sliced mesh representations and profiles of the electron density and Hartree potential are shown in Figure \ref{fig:he-aks-mesh} (lower row). A noticeable difference from the mesh in \Cref{fig:he-aks-mesh} (upper row) is the presence of smaller mesh sizes in regions far away from the origin. As a result, the Hartree potential is better captured, as illustrated in the fourth column of  \Cref{fig:he-aks-mesh} (lower row), from which we can find smoother contour lines than that in \Cref{fig:he-aks-mesh} (upper row) using the first indicator \eqref{eq:indi-ks}.

	While the Hartree potential is approximated accurately and chemical accuracy is obtained, the computational cost increases significantly. The number of mesh grids on the finest mesh reaches 2,874,807, surpassing the number of grids used in the previous section. This increment in the number of mesh grids is primarily caused by the need to capture the variations of the Hartree potential. Such an increase in the computational grid introduces additional complexity and computational overhead, leading to longer computation time and higher memory requirements. This becomes especially problematic when dealing with larger system sizes or when conducting complex simulations. To address this issue and reduce the computational cost, an alternative approach, known as the multi-mesh adaptive method, will be introduced in the next section.

	\section{The multi-mesh adaptive method}
	In the preceding section, we explored the solution of the discretized Kohn--Sham equation on a single mesh, employing either the residual-based \emph{a posteriori} error indicator \eqref{eq:indi-ks} or \eqref{eq:indi-kshar}. The results indicated that employing the first indicator \eqref{eq:indi-ks} alone may fall short of achieving chemical accuracy. On the other hand, the second indicator \eqref{eq:indi-kshar} demonstrated the ability to attain chemical accuracy; however, it came at the cost of significantly increased computational requirements. 
	
	To tackle this challenge, we introduce an approach known as the multi-mesh adaptive method \cite{li2005multi}. The primary objective of this method is to strike a balance between achieving chemical accuracy and managing computational costs effectively. By utilizing multiple meshes instead of a single mesh, the multi-mesh adaptive method offers greater flexibility in adapting the mesh resolution to capture variations for different quantities of interest. 
	
	Specifically, in this method, we will utilize two distinct meshes. Both two meshes will be adapted during the simulation. The first mesh is specifically designed for solving the Kohn--Sham equation, with the primary objective of accurately capturing the variations in the wavefunctions. This mesh is tailored to ensure high resolution in regions where the wavefunctions exhibit significant changes. Conversely, the second mesh is dedicated to solving the Poisson equation \eqref{eq:poisson}. Its primary purpose is to capture the variations in the Hartree potential effectively. This second mesh is strategically designed to provide optimal resolution and precision in regions where the Hartree potential exhibits substantial changes. By employing two separate meshes with specific focus areas, we can ensure that each equation is solved with the appropriate level of accuracy and capture the variations unique to the wavefunctions and the Hartree potential, respectively.
	
	To implement the multi-mesh method, careful handling of two components is crucial. The first component involves effectively managing the mesh grids, allowing for flexible local refinement or coarsening as needed. This management mechanism should also facilitate efficient solution updates from the old mesh to the new mesh. The second component focuses on ensuring efficient and accurate communication between different meshes, particularly in the calculation of integrals. Efficient communication protocols play a vital role in maintaining consistency and accuracy across the various meshes. These requirements can be fulfilled by the hierarchical geometry tree data structure, which will be introduced in detail in the following.
	
	\subsection{The hierarchical geometry tree}
	A well-designed data structure for the mesh grids is needed for an effective management mechanism. In the presented algorithm, the hierarchical geometry tree (HGT) \cite{li2005multi, bao2012h} is utilized. 
	
	Firstly, the mesh structure is described hierarchically, which means that the mesh information is given from the lowest dimension (0-D, the points) to the highest dimension (3-D, the tetrahedron) hierarchically. An element such as a point for 0-D, an edge for 1-D, a triangle for 2-D, a tetrahedron for 3-D is called a geometry. In the hierarchical description of a tetrahedron, all geometries have a belonging-to relationship. For example, if an edge is one of the edges of a triangle, this edge belongs to this triangle. With this hierarchical structure, the geometry information of a tetrahedron can be referred to flexibly, and the refinement and coarsening of a mesh can also be implemented efficiently. 
	
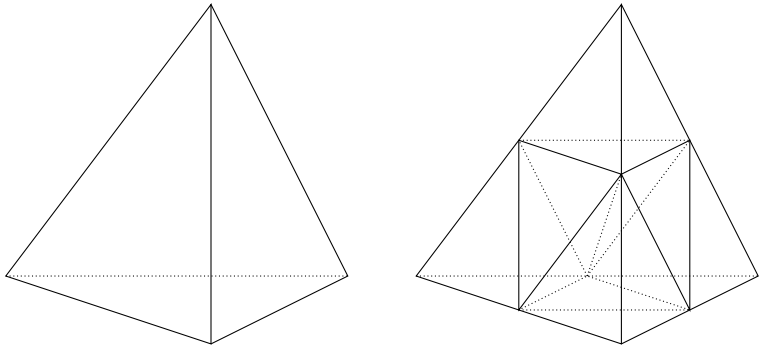
\begin{figure}[h]
		\centering
		\begin{tikzpicture}
			\begin{scope}[scale=0.9]
				\draw[shift={(-6,0)}] (0,0)--(3,-1)--(3,4)--cycle; 
				\draw[shift={(-6,0)}] (3,-1)--(5,0)--(3,4);
				\draw[shift={(-6,0)}, densely dotted] (0,0)--(5,0); 
				
				\draw[] (0,0)--(3,-1)--(3,4)--cycle; 
				\draw[] (3,-1)--(5,0)--(3,4);
				\draw[densely dotted] (0,0)--(5,0); 
				
				\draw[] (1.5,-0.5)--(1.5,2)--(3,1.5)--cycle;
				\draw[] (3,1.5)--(4,2)--(4,-0.5)--cycle;
				\draw[densely dotted] (1.5,-0.5)--(4,-0.5)--(2.5,0)--cycle;
				\draw[densely dotted] (2.5,0)--(1.5,2)--(4,2)--cycle;
				\draw[densely dotted] (2.5,0)--(3,1.5);
				
			\end{scope}
		\end{tikzpicture}
		
		\caption{The parent tetrahedron (left) and the eight child tetrahedrons (right). \label{fig:tet}}
	\end{figure}

	Secondly, the mesh is stored and managed by a tree data structure. The validity of using the tree data structure is due to the strategy of element refinement and coarsening. Specifically, for a tetrahedron element (the left of \Cref{fig:tet}), the refinement of this tetrahedron is dividing it into eight equally small tetrahedrons via connecting the midpoints on each edge (the right of \Cref{fig:tet}). As a result, a belonging-to relationship can be established, for example, any small tetrahedron that is called ``child" belongs to the large ``parent" tetrahedron. Meanwhile, the coarsening of any child tetrahedron in the right of \Cref{fig:tet} is releasing all the children to obtain the parent tetrahedron. By this refinement and coarsening strategy, the tree data structure is able to be established. 
	
\begin{figure}[h]
		\centering
		\begin{tikzpicture}
			\begin{scope}[scale=0.9]
				\draw[shift={(-0.25,-0.1)}] (-3,0) node{A};
				\draw[shift={(0.25,-0.1)}] (1,0) node{B};
				\draw[shift={(0.15,0.2)}] (0,3) node{C};
				\draw (-0.67,1) node {$\mathcal{T}_0$};
				\draw[thick] (-3,0)--(1,0)--(0,3)--cycle;
				\draw[shift={(-0.25,-0.1)}] (3,0) node{A};
				\draw[shift={(0.25,-0.1)}] (7,0) node{B};
				\draw[shift={(0.15,0.2)}] (6,3) node{C};
				\draw[thick] (3,0)--(7,0)--(6,3)--cycle;
				\draw[thick] (4.5,1.5)--(5,0)--(6.5,1.5)--cycle;
				\draw[shift={(-0.21,0.16)}] (4.5,1.5) node{D};
				\draw[shift={(0.06,-0.28)}] (5,0) node{E};
				\draw[shift={(0.25,0.1)}] (6.5,1.5) node{F};
				\draw[shift={(5,-0.5)}] (-0.67,1) node {$\mathcal{T}_{0,0}$};
				\draw[shift={(6,0)}] (-0.67,1) node {$\mathcal{T}_{0,1}$};
				\draw[shift={(7,-0.5)}] (-0.67,1) node {$\mathcal{T}_{0,2}$};
				\draw[shift={(6.4,0.95)}] (-0.67,1) node {$\mathcal{T}_{0,3}$};
				
				\draw[shift={(5.75,-0.1)}] (3,0) node{A};
				\draw[shift={(6.25,-0.1)}] (7,0) node{B};
				\draw[shift={(6.15,0.2)}] (6,3) node{C};
				\draw[thick,shift={(6,0)}] (3,0)--(7,0)--(6,3)--cycle;
				\draw[thick,shift={(6,0)}] (4.5,1.5)--(5,0)--(6.5,1.5)--cycle;
				\draw[shift={(5.79,0.16)}] (4.5,1.5) node{D};
				\draw[shift={(6.06,-0.28)}] (5,0) node{E};
				\draw[shift={(6.25,0.1)}] (6.5,1.5) node{F};
				\draw[thick,shift={(6,0)}]  (3.75,0.75)--(4,0)--(4.75,0.75)--cycle;
				\draw[shift={(5.79,0.16)}] (3.75,0.75) node{G};
				\draw[shift={(6.06,-0.28)}] (4,0) node{H};
				\draw[shift={(6.25,0.1)}] (4.75,0.75) node{I};
				\draw (9.60,0.15) node {{\tiny$\mathcal{T}_{0,0,0}$}};
				\draw (10.1,0.5) node {{\tiny$\mathcal{T}_{0,0,1}$}};
				\draw (10.6,0.15) node {{\tiny$\mathcal{T}_{0,0,2}$}};		
				\draw (10.3,0.9) node {{\tiny$\mathcal{T}_{0,0,3}$}};	
				\draw[shift={(12.2,0)}] (-0.67,1) node {$\mathcal{T}_{0,1}$};
				\draw[shift={(13,-0.5)}] (-0.67,1) node {$\mathcal{T}_{0,2}$};
				\draw[shift={(12.5,0.95)}] (-0.67,1) node {$\mathcal{T}_{0,3}$};
				
			\end{scope}		
		\end{tikzpicture}
		\caption{The refinement of a triangle $\mathcal{T}_0$ and its sub-triangle $\mathcal{T}_{0,0}$. \label{fig:refine}}
	\end{figure}
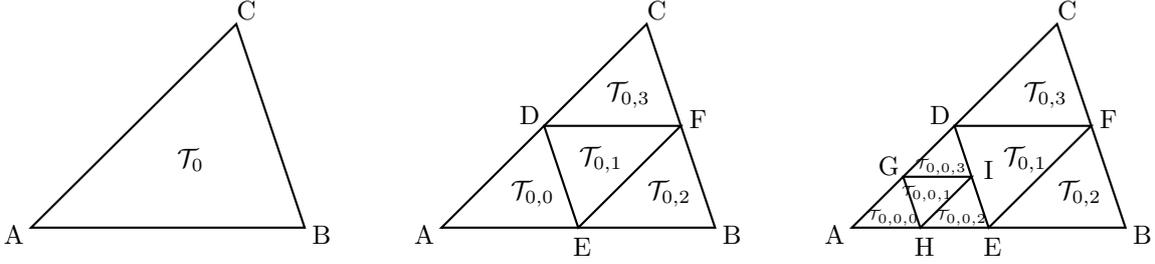

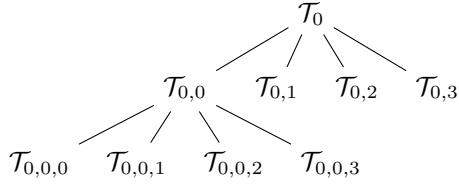
\begin{figure}[h]
		\centering
		\begin{forest}
			[$\mathcal{T}_{0}$
			[$\mathcal{T}_{0,0}$
			[$\mathcal{T}_{0,0,0}$]
			[$\mathcal{T}_{0,0,1}$]
			[$\mathcal{T}_{0,0,2}$]
			[$\mathcal{T}_{0,0,3}$]
			]
			[$\mathcal{T}_{0,1}$]
			[$\mathcal{T}_{0,2}$]
			[$\mathcal{T}_{0,3}$]
			]
		\end{forest}
		\caption{The quadtree data structure for the mesh in the right of \Cref{fig:refine}. \label{fig:quadtree}}
	\end{figure}

	For a better understanding of the tree data structure, a two-dimensional illustration is presented. Similar to the 3D case, the refinement of a 2D element, i.e. the triangle, is to divide the triangle into four equal triangles, as displayed in the left two columns in \Cref{fig:refine}. By refining a triangle $\mathcal{T}_0$, four sub-triangles $\{\mathcal{T}_{0,0},\mathcal{T}_{0,1},\mathcal{T}_{0,2},\mathcal{T}_{0,3}\}$ are generated. Furthermore, via refining $\mathcal{T}_{0,0}$ the sub-triangles $\{\mathcal{T}_{0,0,0},\mathcal{T}_{0,0,1},\mathcal{T}_{0,0,2},\mathcal{T}_{0,0,3}\}$  are obtained, which is demonstrated in \Cref{fig:refine}. To manage this procedure. the quadtree data structure in which each internal node has exactly four children is utilized, as described in \Cref{fig:quadtree}. When the local refinement and coarsening techniques are adopted, only some triangles are to refined, as indicated in the right of \Cref{fig:refine} and the bottom of \Cref{fig:quadtree} where only the triangle $\mathcal{T}_{0,0}$ is refined. In the quadtree, we call $\mathcal{T}_0$ the root node, and those nodes without further descents like $\mathcal{T}_{0,1}$ and $\mathcal{T}_{0,0,1}$ the leaf nodes. Suppose that there is a set of root nodes $\{\mathcal{T}_i\}, i=0,1,2,\dots$, which form a mesh for a domain $\Omega$, then from the above definition we know that a set of all leaf nodes also form a mesh. For example, the set $\{\mathcal{T}_0\}$ forms the mesh in the left of \Cref{fig:refine}, the set  $\{\mathcal{T}_{0,0},\mathcal{T}_{0,1},\mathcal{T}_{0,2},\mathcal{T}_{0,3}\}$ forms the middle mesh in \Cref{fig:refine}, and the set $\{\mathcal{T}_{0,0,0},\mathcal{T}_{0,0,1},\mathcal{T}_{0,0,2},\mathcal{T}_{0,0,3},\mathcal{T}_{0,1},\mathcal{T}_{0,2},\mathcal{T}_{0,3}\}$ forms the right mesh in \Cref{fig:refine}.
	
\begin{figure}[!h]
		\centering
		\begin{tikzpicture}
			\begin{scope}
				\path (0,0) coordinate (D) node [below left] {$D$};
				\path (4,0) coordinate (E) node [below right] {$E$};
				\path (2,0) coordinate (I) node [below] {$I$};			
				\path (3,3) coordinate (F) node [above] {$F$};			 
				\draw[thick] (D) -- (E) -- (F) -- cycle ;
				\draw[dashed] (F) -- (I);
				
				\draw[fill=black] (D) circle (1.5pt) (E) circle (1.5pt) (F) circle (1.5pt) (I) circle (1.5pt);
			\end{scope}
		\end{tikzpicture}
		\caption{Twin-triangle. \label{fig:twin-tri}}
	\end{figure}
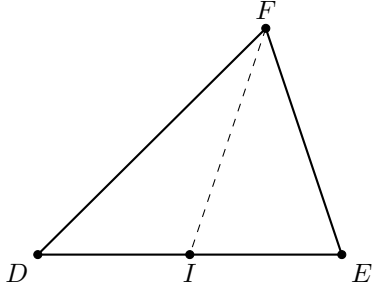

	With the hierarchical description of the geometry and the tree data structure, the mesh is effectively managed by the HGT. However, building the finite element space directly on a mesh results in non-conforming finite element because of the hanging points in the direct neighbors of those refined triangles. These hanging points can be handled mainly in two ways. If there are more than one hanging point on a triangle, this triangle will be refined. This process is called the semi-regularization procedure \cite{li2005multi}. If there exists only one hanging point on a triangle,  for example, the triangle $\triangle DEF$ in the right of \Cref{fig:refine} has a hanging point $I$ arose in the refinement of $\mathcal{T}_{0,0}$. To deal with these kind of hanging points, the twin-triangle geometry is introduced, which is demonstrated in \Cref{fig:twin-tri}. The twin-triangle actually consists of two triangles $\triangle FDI$ and $\triangle FIE$. Meanwhile, the degrees of freedom and the interpolation information of the twin-triangle $\triangle DEF$ inherit from the two triangles $\triangle FDI$ and $\triangle FIE$. To make the finite element space conforming, the following strategy is adopted to construct the basis functions in the twin-triangle geometry: For each basis function, its value is 1 at its corresponding interpolation point and 0 at other interpolation points. The support of the basis function whose interpolation point is a common point of two sub-triangles in a twin-triangle geometry such as $F$ and $I$ in \Cref{fig:twin-tri}, is the whole twin-triangle geometry. For the non-common points like $D$, the support of its basis function is only the triangle   $\triangle FDI$. With such a strategy, a conforming finite element space can be built in a mesh where the local refinement is implemented. 
	
	In virtue of the tree data structure, multiple meshes are allowed to be described by the same HGT. Naturally, a problem arises: is it possible to make the information communicates among these meshes without any loss? The answer is affirmative, and the reason relies on the belonging-to relationship between any two nodes in the HGT. Briefly speaking, there exist only three kinds of relationship between any two nodes: equal, belonging-to, and no-overlap. The communication among equal or no-overlap elements is trivial. Consequently, we only need to take care of the second kind of relationship. We introduce the implementation details in the following.
	
\begin{figure}[!h]
		\centering
		\begin{tikzpicture}
			\begin{scope}
				\path (0,0) coordinate (A) node [below left] {$A$};
				\path (4,0) coordinate (B) node [below right] {$B$};		
				\path (3,3) coordinate (C) node [above] {$C$};			 
				\path (1.5,1.5) coordinate (D) node [left] {$D$};
				\path (2,0) coordinate (E) node [below] {$E$};	
				\path (3.5,1.5) coordinate (F) node [right] {$F$};
				\path (1.75,0.75) coordinate (I) node [above right] {$I$};
				\path (1,0.) coordinate (H) node [below] {$H$};
				\path (0.75,0.75) coordinate (G) node [left] {$G$};
				
				\draw[thick] (A) -- (B) -- (C) -- cycle ;
				\draw[thick] (D) -- (E) -- (F) -- cycle ;
				\draw[thick] (G) -- (H) -- (I) -- cycle ;
				\draw[dashed] (F) -- (I);
				
				\draw[fill=black] (A) circle (1.5pt) (B) circle (1.5pt) (C) circle (1.5pt) (D) circle (1.5pt) (E) circle (1.5pt) (F) circle (1.5pt) (G) circle (1.5pt) (H) circle (1.5pt) (I) circle (1.5pt);
				
				\path (6,0) coordinate (A1) node [below left] {$A$};
				\path (10,0) coordinate (B1) node [below right] {$B$};		
				\path (9,3) coordinate (C1) node [above] {$C$};			 
				\path (7.5,1.5) coordinate (D1) node [left] {$D$};
				\path (8,0) coordinate (E1) node [below] {$E$};	
				\path (9.5,1.5) coordinate (F1) node [right] {$F$};
				\path (8.75,0.75) coordinate (J) node [above] {$J$};
				\path (9,0.) coordinate (K) node [below] {$K$};
				\path (9.75,0.75) coordinate (L) node [right] {$L$};
				
				\draw[thick] (A1) -- (B1) -- (C1) -- cycle ;
				\draw[thick] (D1) -- (E1) -- (F1) -- cycle ;
				\draw[thick] (J) -- (K) -- (L) -- cycle ;
				\draw[dashed] (D1) -- (J);
				
				\draw[fill=black] (A1) circle (1.5pt) (B1) circle (1.5pt) (C1) circle (1.5pt) (D1) circle (1.5pt) (E1) circle (1.5pt) (F1) circle (1.5pt) (K) circle (1.5pt) (J) circle (1.5pt) (L) circle (1.5pt);

			\end{scope}
		\end{tikzpicture}
		\caption{Left: $\mathcal{T}_{\mathrm{KS}}$. Right:  $\mathcal{T}_{\mathrm{Har}}$. \label{fig:two-tri-mesh}}
	\end{figure}
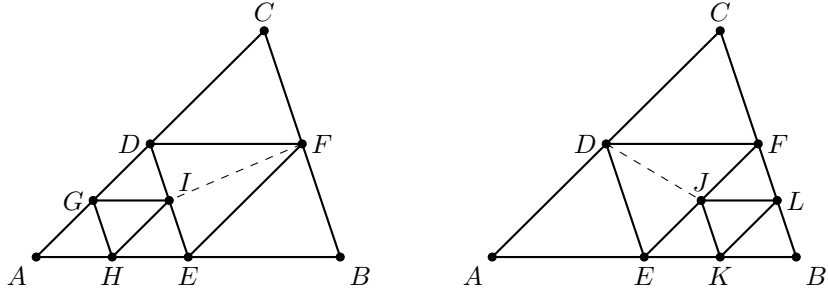

	In the presented method, the Hamiltonian matrix and energy require information from both the KS mesh and the Hartree mesh, which mainly involve numerical integration. We take the evaluation of the Hartree potential energy for example, which could be written as 
	\begin{equation}\label{eq:har-cal}
		E_{\mathrm{Har}} = \frac{1}{2}\int_\Omega \phi(\mathbf{r}) \left(\sum_{l=1}^p\psi_l(\mathbf{r})^2 \right)\dr. 
	\end{equation}
	The integration should be carefully treated since the Hartree potential  $\phi(\mathbf{r})$ and the wavefunction $\psi_l(\mathbf{r})$ belong to different finite element spaces built on different meshes. An intuitive illustration is given to show how the presented method calculates the integral \eqref{eq:har-cal}. Assume the finite element space $V_{\mathcal{T}_{\mathrm{KS}}}$ for the wavefunction is built on the mesh $\mathcal{T}_{\mathrm{KS}}$, and the space $V_{\mathcal{T}_{\mathrm{Har}}}$ is built on the mesh $V_{\mathcal{T}_{\mathrm{Har}}}$, see \Cref{fig:two-tri-mesh}. For the common elements such as $\triangle CDF$ which belongs to both $\mathcal{T}_{\mathrm{KS}}$ and $\mathcal{T}_{\mathrm{Har}}$,  the numerical integration can be calculated directly.  While for the remained elements, for example, in $\mathcal{T}_{\mathrm{KS}}$, the triangle $\triangle DAE$ is refined, while in $\mathcal{T}_{\mathrm{Har}}$ this triangle is kept, and a similar case for the triangle $\triangle FEB$, special treatment is needed to avoid the loss of accuracy. 
	
	In order to prevent the loss of accuracy, we employ a strategy that maximizes the utilization of quadrature points in numerical integrals. Specifically, the numerical integration on the element $\triangle ADE$ is divided into the integrations on its four refined sub-triangles, 
	\begin{align*}
		\frac{1}{2}\int_{\triangle ADE} \phi(\mathbf{r}) \left(\sum_{l=1}^p \psi_l(\mathbf{f}) ^2 \right) \dr \approx \sum_{K=1}^4\sum_{j=1}^q \mbox{area}(K) J_j^K w_j^K \phi(\mathbf{r}_j^K)\left(\sum_{l=1}^p \psi_l(\mathbf{r}_j^K)^2 \right),
	\end{align*}
	where the element $K$ represent the four sub-triangles of $\triangle ADE$, $\mathbf{r}_j^K$ is the $j$-th quadrature point of $K$, $J_j^K$ is the jacobian at $\mathbf{r}_j^K$, and $w_l^K$ is the associated weight in the numerical quadrature. The values of the Hartree potential on these quadrature points can be obtained by numerical interpolation. Similarly, the integral on the triangle $\triangle FEB$ is evaluated by doing the numerical integral on its sub-triangles. In this way, the accuracy of the integral will not be affected in the communication of different meshes.

	With the HGT, the solution update from the old mesh to the new mesh after the local refinement and coarsening can also be implemented efficiently. As we mentioned before, a set of all leaf nodes forms a mesh. Although different meshes correspond to different sets of leaf nodes, all meshes are from the same set of root nodes. Then the belonging-to relationship of the geometries between two meshes can be analyzed easily. As a result, the solution update can be implemented efficiently according to the relationship.

	\subsection{The multi-mesh adaptive algorithm}
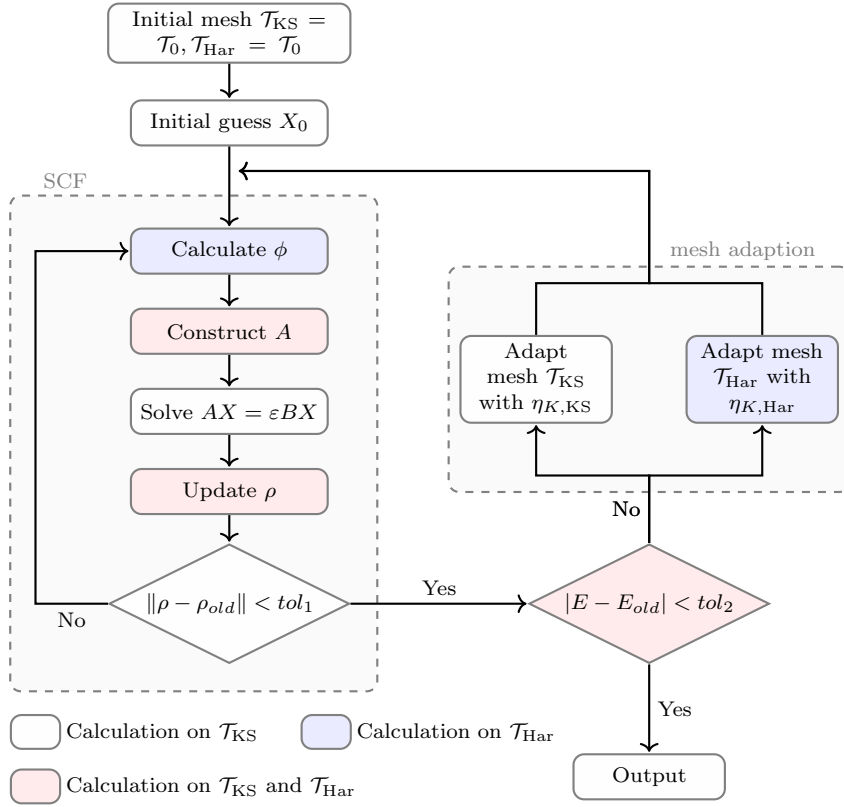
\begin{figure}[h]
		\centering \footnotesize
		\begin{tikzpicture}[
			scale=.85,
			auto,
			decision/.style = { diamond, aspect=2, draw=gray,
				thick, fill=gray!1, text width=4em, text badly centered,
				inner sep=1pt},
			block/.style = { rectangle, draw=gray, thick, fill=gray!1,
				text width=6em, text centered, rounded corners,
				minimum height=2em },
			blockhar/.style = { rectangle, draw=gray, thick, fill=blue!8,
				text width=8em, text centered, rounded corners,
				minimum height=2em },
			blockkshar/.style = { rectangle, draw=gray, thick, fill=red!8,
				text width=8em, text centered, rounded corners,
				minimum height=2em },
			line/.style = { draw, thick, ->, shorten >= 0.5pt},
			]
			
			\node [block, text width=10em] at (0,1.9) (initmesh) {Initial mesh $\mathcal{T}_{\mathrm{KS}} = \mathcal{T}_0, \mathcal{T}_{\mathrm{Har}} = \mathcal{T}_0$ };
			\node [block, text width=8em] at (0,0.5) (prob) {Initial guess $X_0$};

			\node at (0,-0.25) (null1) {};
			
			\node [block, text width=1.5em,minimum height=1.5em] at (-3,-9) (harmesh) {};
			\node [right] at (-2.65,-9) {Calculation on $\mathcal{T}_{\mathrm{KS}}$};
			\node [blockhar, text width=1.5em,minimum height=1.5em] at (1.5,-9) (ksmesh) {};
			\node [right] at (1.85,-9) {Calculation on $\mathcal{T}_{\mathrm{Har}}$};
			\node [blockkshar, text width=1.5em,minimum height=1.5em] at (-3,-9.85) (ksmesh) {};
			\node [right] at (-2.65,-9.85) {Calculation on $\mathcal{T}_{\mathrm{KS}}$ and $\mathcal{T}_{\mathrm{Har}}$ };
			
			\node [block,rectangle, draw=gray, thick,dashed, fill=gray!4,
			text width=15.5em, text centered, rounded corners, minimum
			height=22em] at (-0.55,-4.5) (scf) {};
			\node [right, color=gray] at (-3,-0.4) {SCF};
			
			\node [block,rectangle, draw=gray, thick,dashed, fill=gray!4,
			text width=17em, text centered, rounded corners, minimum
			height=10em] at (6.5,-3.5) (ada) {};
			\node [right,color=gray] at (6.7,-1.5) {mesh adaption};

			\node [blockhar] at (0,-1.5) (genhar) {Calculate $\phi$};
			
			\node [blockkshar] at (0,-2.75) (genh) {Construct $A$};
			\node [block, text width = 8em] at (0,-4)  (eigensolver) {Solve $AX = \varepsilon BX$};
			\node [blockkshar] at (0, -5.25) (updaterho) {Update $\rho$};
			\node [decision, text width=8em] at (0,-7) (scfdec) {$\|\rho-\rho_{old}\| < tol_1 $};
			
			\node [decision, text width=8em,fill=red!8] at (6.5,-7) (adadec) {$|E-E_{old}| < tol_2$};
			
			\node [block,text width=6em] at (4.75,-3.5) (adaks) {Adapt mesh $\mathcal{T}_{\mathrm{KS}}$ with $\eta_{K,\mathrm{KS}}$};
			
			\node [blockhar,text width=6em] at (8.25,-3.5) (adahar) {Adapt mesh $\mathcal{T}_{\mathrm{Har}}$ with $\eta_{K,\mathrm{Har}}$};
			
			\node [block] at (6.5, -9.7) (end) {Output};

			\begin{scope} [every path/.style=line,thick,shorten >= 0.5pt]
				\path (initmesh) -- (prob);
				\path (prob)    --   (genhar);   
				\path (genhar)    --   (genh);
				\path (genh) -- (eigensolver);
				\path (eigensolver) -- (updaterho);
				\path (updaterho) -- (scfdec);
				\path (scfdec) -- node{No} (-3,-7)-- (-3,-1.5) -- (genhar);
				\path (scfdec) -- node{Yes} (adadec);
				\path (adadec) -- node{No}(6.5,-5)-| (adaks);
				\path (adadec) -- node{No}(6.5,-5)-| (adahar);
				\path (adahar) |- (6.5,-2) |- (null1);
				\path (adaks) |- (6.5,-2) |- (null1);
				\path (adadec) -- node{Yes} (end);
			\end{scope}
			
		\end{tikzpicture}
		\caption{Flowchart of the multi-mesh adaptive algorithm for the KS equation. \label{fig:algorithm}}
	\end{figure}

	In virtue of the HGT data structure, the multi-mesh method can be efficiently implemented. To solve the Kohn--Sham equation, we propose the multi-mesh adaptive algorithm illustrated in the flowchart presented in \Cref{fig:algorithm}. The Kohn--Sham equation \eqref{eq:KS} is discretized and solved on the finite element space built on the mesh $\mathcal{T}_{\mathrm{KS}}$, and the Hartree potential \eqref{eq:poisson} is solved on the mesh $\mathcal{T}_{\mathrm{Har}}$. The KS mesh $\mathcal{T}_{\mathrm{KS}}$ is adapted using the error indicator \eqref{eq:indi-ks} and the Har mesh $\mathcal{T}_{\mathrm{Har}}$ is adapted using the error indicator \eqref{eq:indi-poisson}. Both of these two mesh adaptions occur after the end of the SCF iteration. The mesh adaptations continue until the final total energy reaches convergence, at which point the mesh adaptation ceases, and the results are outputted.

	In order to assess the efficiency and make a comparison between the multi-mesh algorithm \Cref{fig:algorithm} and the single mesh algorithm \Cref{alg:single}, we categorize the calculations into three distinct types. Namely, the calculations on the Kohn--Sham mesh $\mathcal{T}_{\mathrm{KS}}$ with a white background in \Cref{fig:algorithm}, the calculations conducted in the Hartree mesh $\mathcal{T}_{\mathrm{Har}}$ with a blue background, and the calculations that require information from both $\mathcal{T}_{\mathrm{KS}}$ and$\mathcal{T}_{\mathrm{Har}}$ with a red background.  When comparing the multi-mesh algorithm to the single mesh algorithm, it becomes evident that additional calculations fall into the third category. These calculations necessitate communication between the Kohn--Sham mesh and the Hartree mesh. Specifically, such communication is required during the generation of the Hamiltonian, the updating of the electron density (which forms the right-hand side of the Poisson equation \eqref{eq:poisson} for the Hartree potential), and the computation of the total energy. We will conduct a detailed comparison of the time consumed by these individual parts to evaluate the overall performance of the algorithms in the next section.

	By employing the multi-mesh method and incorporating adaptive algorithms guided by appropriate error indicators, we can achieve accurate and efficient solutions for the Kohn--Sham equation. The flexibility of mesh adaptation ensures that the mesh resolution aligns with the specific requirements of each equation, ultimately leading to improved convergence and reliable results. This can be verified by the Helium example.

	\section{Numerical Experiments}\label{sec:numEx}
	In this section, we examine the convergence and efficiency of the multi-mesh adaptive method through a series of numerical examples. All the simulations are performed on a workstation ``Moss" with two AMD EPYC 7713 64-Core Processors (at 2.0GHz$\times$64, 512M cache) and 900GB of RAM, and the total number of cores is 128. The software is the C++ library \texttt{AFEABIC} \cite{bao2012h} under Ubuntu 20.04.
	\subsection{Case study: for chemical accuracy and computational efficiency}
	The multi-mesh adaptive method \Cref{fig:algorithm} is first tested for three cases: the helium atom, the LiH molecule, and the \ch{BeH2} molecule. Our examination unfolds in three phases: firstly, we assess the method's systematic convergence in the context of the helium atom example; next, we determine whether it can attain chemical accuracy across all three cases; and lastly, we undertake a comparative analysis of the computational costs required to achieve chemical accuracy between the multi-mesh method and the single-mesh method.
	\subsubsection{Helium atom}
\begin{figure}[!h]
		\centering
		\includegraphics[width=0.45\linewidth]{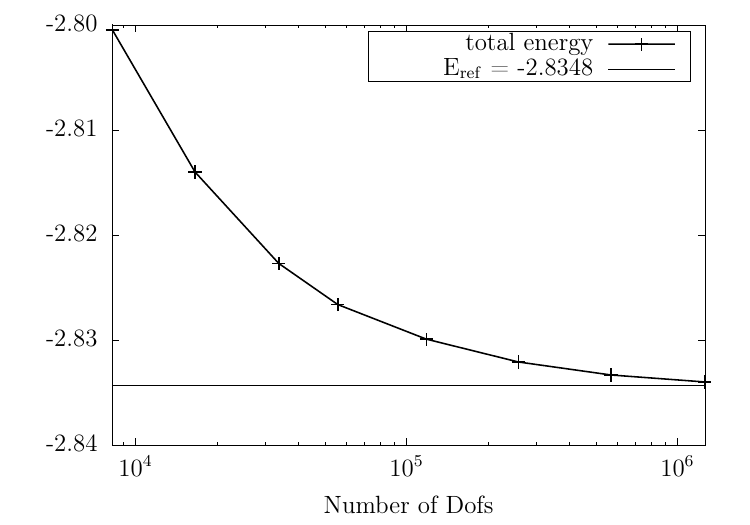}~
		\includegraphics[width=0.45\linewidth]{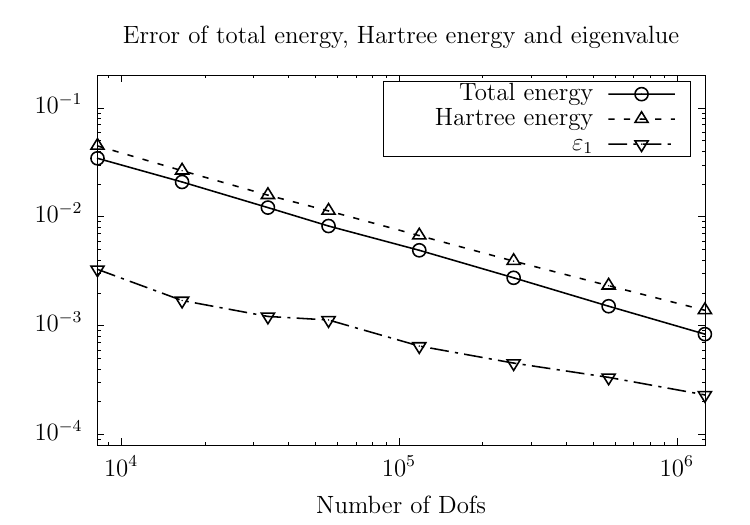}
		\caption{Convergence history of the total energy (left), absolute errors of energies, and eigenvalue (right) for the multi-mesh adaptive method.\label{fig:he-aksahar}}
	\end{figure}

	Similar to the adaptive method using error indicator \eqref{eq:indi-kshar}, the systematical convergence of the energies and eigenvalue are observed in \Cref{fig:he-aksahar}, and the chemical accuracy is obtained when the number of degrees of freedom in $\mathcal{T}_{\mathrm{KS}}$ achieves 1,262,599, which is far less than that requires in the single-mesh algorithm. Furthermore, there are 2,185,920 degrees of freedom in the Hartree mesh $\mathcal{T}_{\mathrm{KS}}$, which is also less than the number of mesh grids used to achieve the chemical accuracy using the single mesh adaptive method. 
	
	\begin{figure}[!h]
		\centering
		\includegraphics[width=0.3\linewidth]{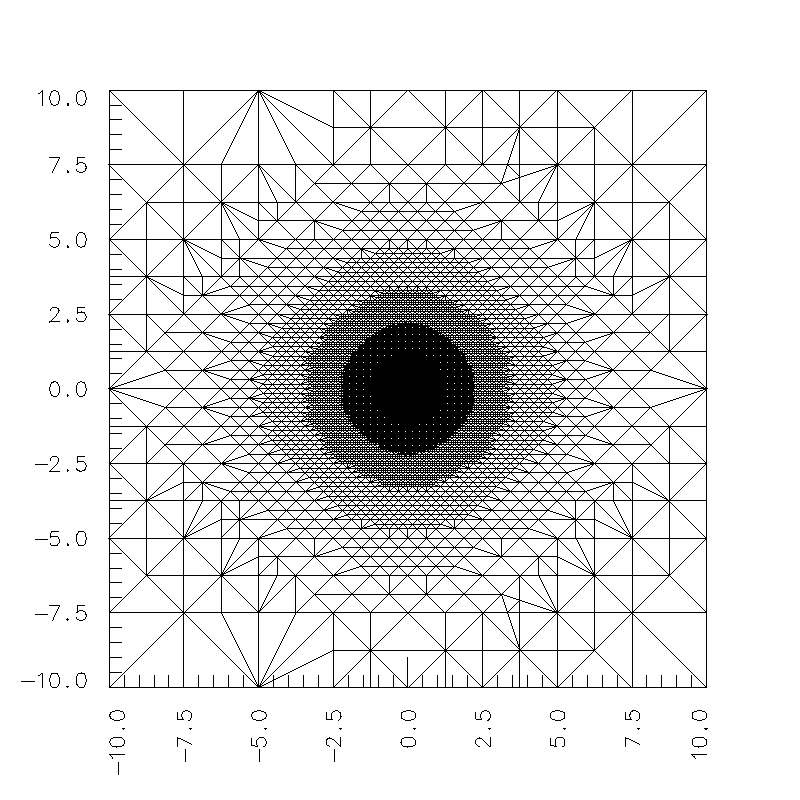}
		\includegraphics[width=0.3\linewidth]{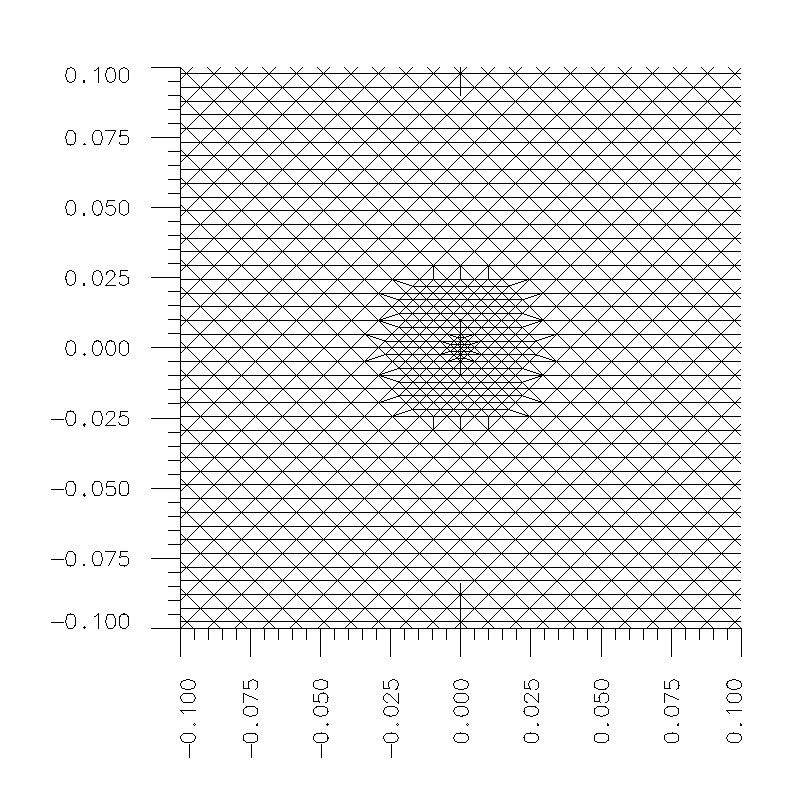}
		\includegraphics[width=0.3\linewidth]{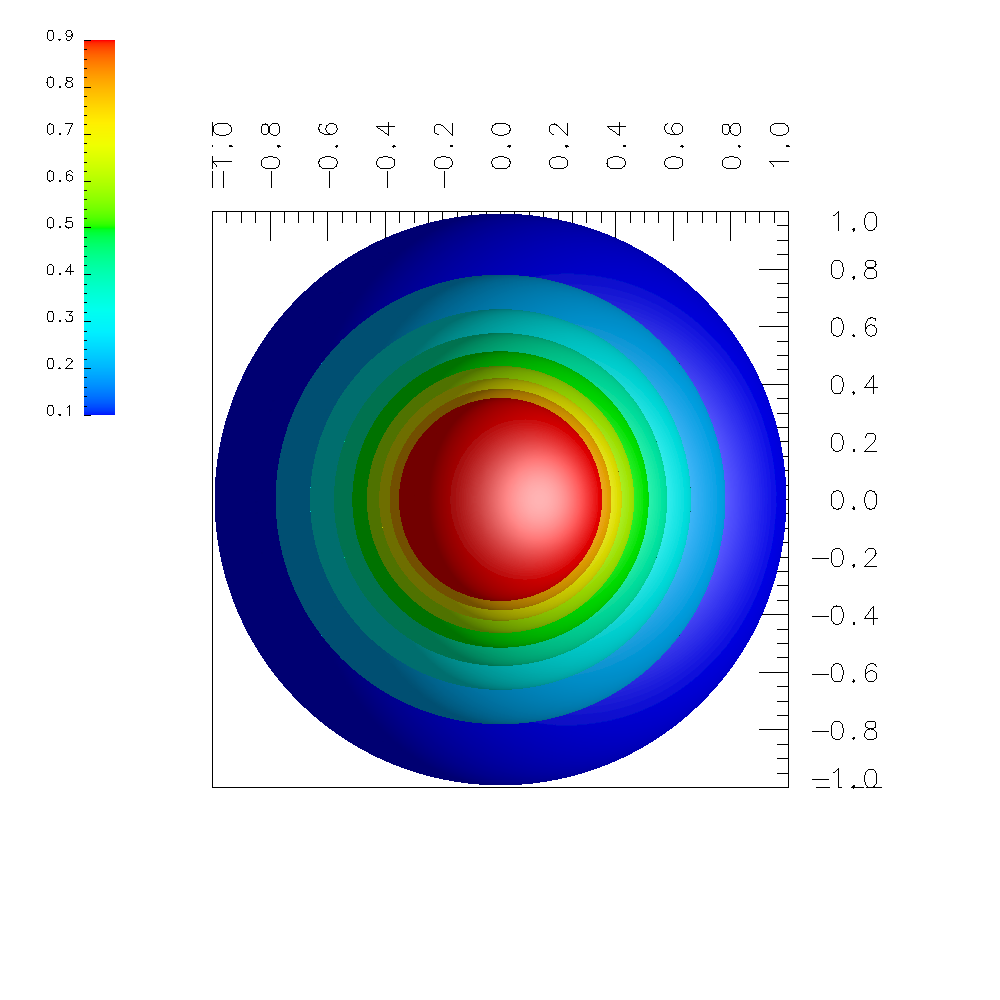}\\
		\includegraphics[width=0.3\linewidth]{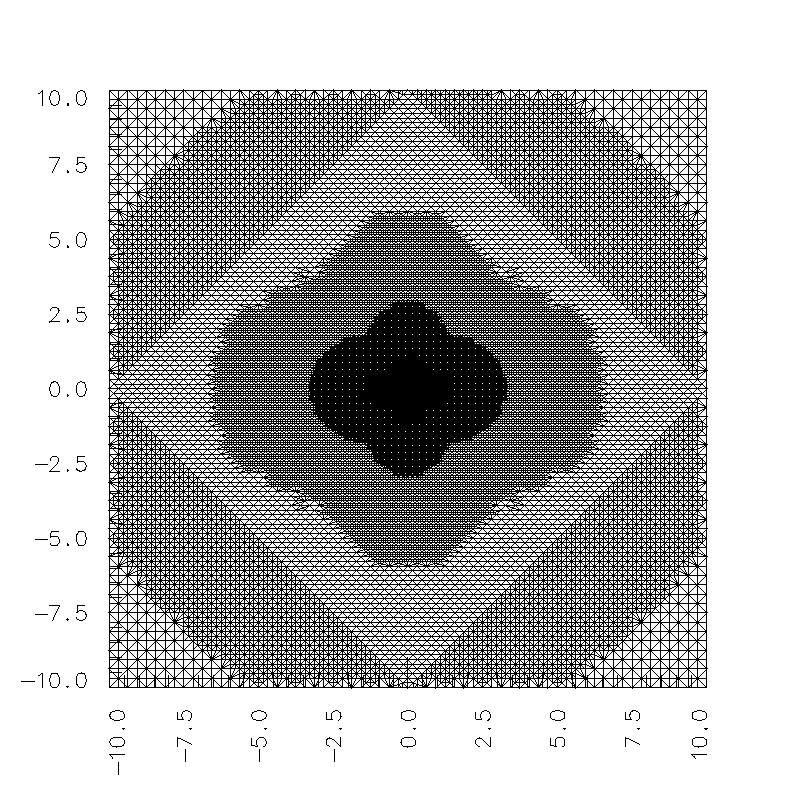}
		\includegraphics[width=0.3\linewidth]{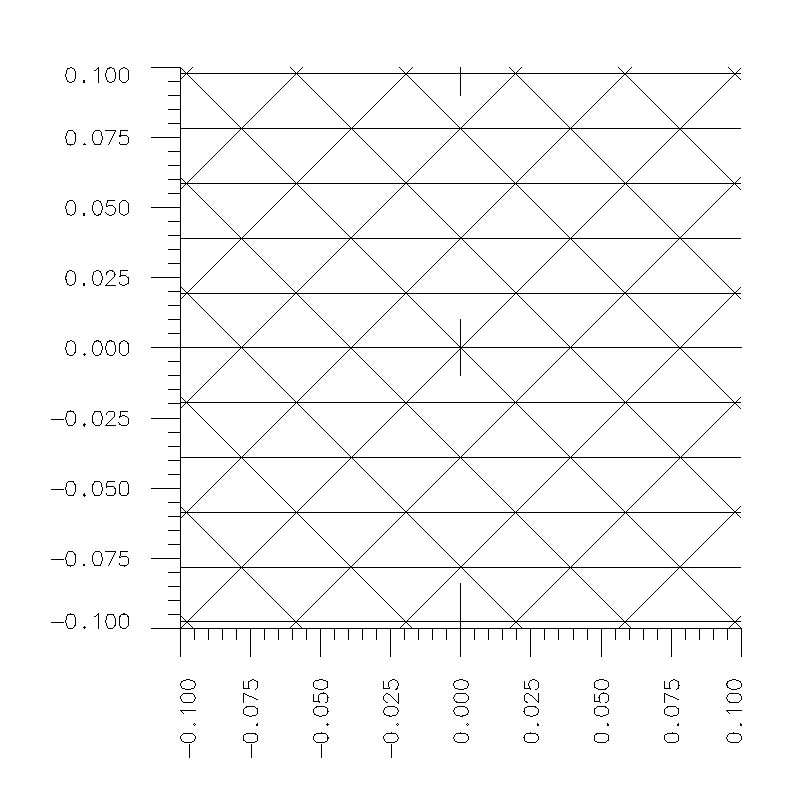}
		\includegraphics[width=0.3\linewidth]{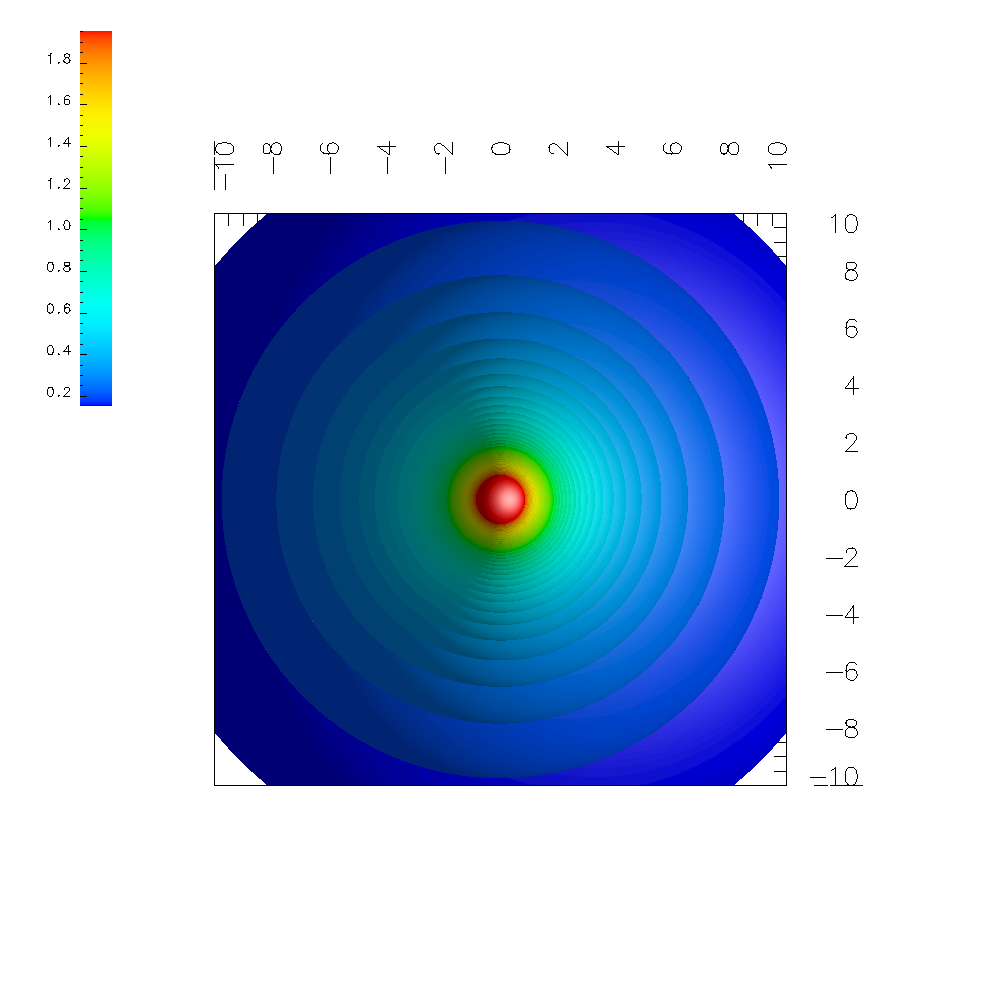}
		\caption{Global and zoomed-in meshes on the sliced $X$-$Y$ plane (top: $\mathcal{T}_\mathrm{KS}$ (1,2262,599 mesh grids); bottom: $\mathcal{T}_\mathrm{Har}$ (2,185,920 mesh grids)), density profile (top right), and Hartree potential profile (bottom right) for the multi-mesh adaptive method. \label{fig:he-aksahar-mesh}}
	\end{figure}
	
	The final meshes are displayed in \Cref{fig:he-aksahar-mesh}. In the top of  \Cref{fig:he-aksahar-mesh} is the sliced mesh of $\mathcal{T}_{\mathrm{KS}}$ which is quite similar with \Cref{fig:he-aks-mesh} (upper row). The meshes of $\mathcal{T}_\mathrm{Har}$ are demonstrated in the bottom of \Cref{fig:he-aksahar-mesh}, which has a similar mesh grid distribution as that in \Cref{fig:he-aks-mesh} (lower row) as shown in a global vision, while it has a sparser grid distribution comparing with \Cref{fig:he-aks-mesh} (lower row) since the Hartree potential behaves less singular than the Kohn--Sham wavefunctions. Although there is a larger number of mesh grids in $\mathcal{T}_\mathrm{Har}$ than in $\mathcal{T}_\mathrm{KS}$, it should be mentioned that on $\mathcal{T}_\mathrm{Har}$ we only need to solve a linear system, while in $\mathcal{T}_\mathrm{KS}$, a nonlinear eigenvalue problem should be solved. Consequently, the major computational cost is from the calculations on $\mathcal{T}_\mathrm{KS}$, and the increment of the computational cost due to the large number of mesh grids on $\mathcal{T}_\mathrm{Har}$ is slight. As a result, the computational cost of the multi-mesh method is of the same order of magnitude of the adaptive method,  while it can achieve chemical accuracy. 
	
\begin{figure}[!h]
		\centering
		\includegraphics[width=0.45\linewidth]{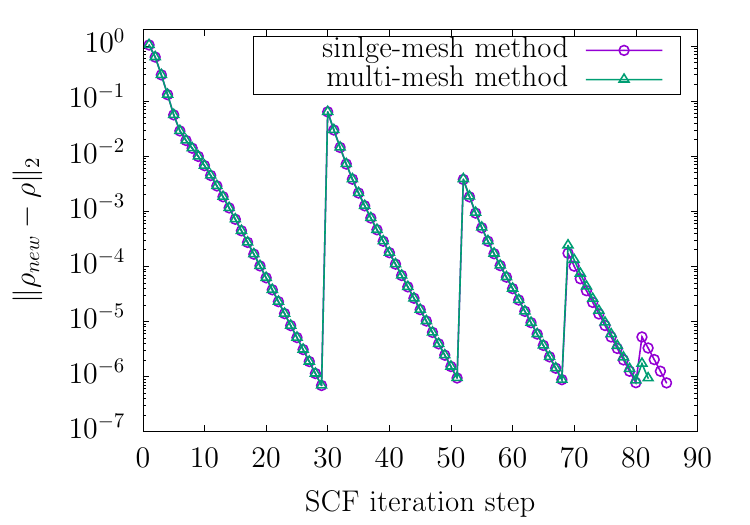}
		$~$
		\includegraphics[width=0.45\linewidth]{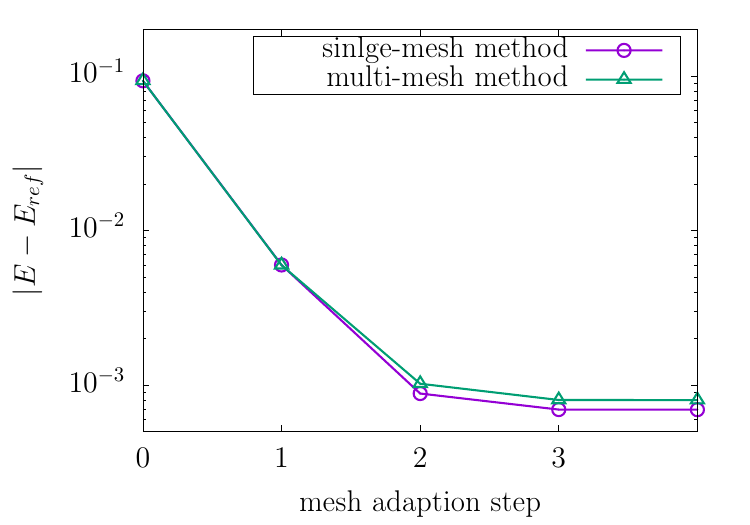}
		\caption{SCF convergence history (left) and mesh adaption history (right) for Helium. \label{fig:hehis}}
	\end{figure}

	As discussed in previous sections, both the single-mesh and multi-mesh methods are able to achieve chemical accuracy when the Hartree potential is considered in constructing the error indicator. To demonstrate the efficiency of the multi-mesh method, we compare the serial time for these two methods to achieve chemical accuracy. For the sake of fairness, we start the two algorithms from the same mesh and initial guess, and the tolerance for guiding the mesh adaption is also the same. With these settings, it is found that the convergence in SCF is quite similar and the number of mesh adaptions is the same, as displayed in \Cref{fig:hehis}. From the figure, it suggests that the introduction of Hartree mesh will not affect the convergence of the SCF procedure in KS mesh. Therefore, the comparison of the CPU time is fair and effective.
	
	A summary of the comparison is illustrated in \Cref{tab:heTime}. Both methods achieved chemical accuracy, and the number of mesh grids in the single-mesh method is the largest among all meshes. The CPU time is compared with respect to two parts: the SCF part and the mesh adaption part. From \Cref{tab:heTime}, the multi-mesh method is proved to be more efficient than the single-mesh method in achieving the chemical accuracy of the helium atom.
	\begin{table}[!h]
		\caption{Comparison on \ch{He} with respect to results and serial computational time. $t_{all}$ represents the total CPU time. $t_{SCF}$ stands for the time for SCF and  $t_{MA}$ represents the time for mesh adaption. \label{tab:heTime}}
		\centering
		\begin{tabular}{ rrrrrrrr}
			\toprule
			&$E_{tot}$ &$\Delta E$ &$N_{KS}$  &$N_{Har}$& $t_{all}$ & $t_{SCF}$ &$t_{MA}$\\
			\midrule
			single-mesh&-2.8341&0.0007 &2,874,807  & 2,874,807 &10290.45 &7585.62 &2704.83\\\midrule
			multi-mesh&-2.8340&0.0008 &1,262,599 &2,185,920&7063.97 & 5135.69 &1928.28\\
			\bottomrule  
		\end{tabular}
		
	\end{table}
	
\begin{figure}[h]
		\centering
		\includegraphics[width=\linewidth]{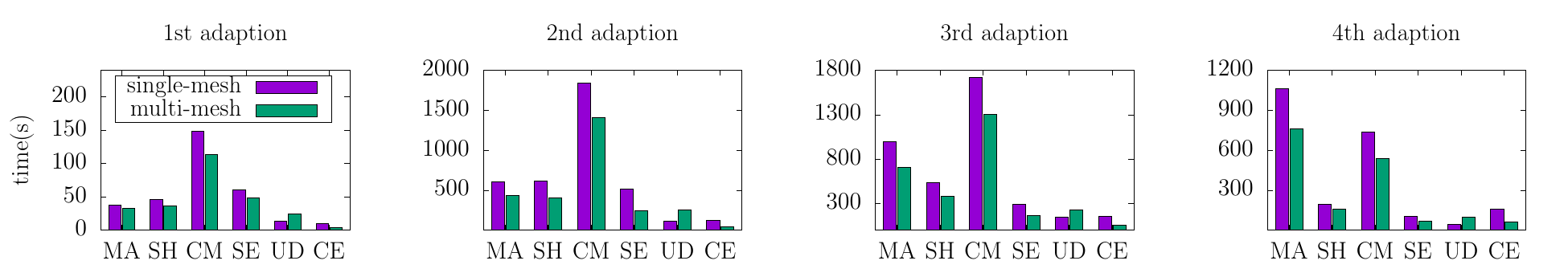}
		\caption{CPU time for single-mesh method and multi-mesh method on six parts:  \textbf{MA} (mesh adaption), \textbf{SH} (solve Hartree potential), \textbf{CM} (construct matrix), \textbf{SE} (solve eigenvalue problem), \textbf{UD} (update electron density) and \textbf{CE} (calculate energy). Four mesh adaptations are needed to achieve chemical accuracy in both methods starting from the same initial setup. \label{fig:hetime}}
	\end{figure}

	A more detailed comparison is presented in \Cref{fig:hetime}. The method can be divided into six parts as indicated in \Cref{fig:hetime}. The four images represent CPU time on four meshes during the mesh adaption process.  Obviously, the cost for solving the Hartree potential and the eigenvalue value problem is less in the multi-mesh method, as expected. Furthermore, the cost of constructing a matrix in the multi-mesh method is also less than that in the single-mesh method on the finest mesh. It is noted that this part contributes the largest portion of the total cost since in the helium example the eigensolver only needs to solve one eigenpair. In larger systems, solving the eigenvalue problem will be the most time-consuming part. 
	
	\subsubsection{LiH molecule}
	A similar comparison for the LiH molecule is employed. The referenced total energy is $E_\mathrm{ref}=-7.9195$ Hartree. The results are displayed in \Cref{tab:LiHTime}, revealing the attainment of chemical accuracy. From this table, the multi-mesh approach demonstrates its efficiency compared to the single-mesh method.
	\begin{table}[!h]
		\caption{Comparison on \ch{LiH} with respect to results and serial computational time. The referenced total energy is $E_\mathrm{ref}=-7.9195$ Hartree. \label{tab:LiHTime}}
		\centering
		\begin{tabular}{ rrrrrrrr}
			\toprule
			&$E_{tot}$ &$\Delta E/2$ &$N_{KS}$  &$N_{Har}$& $t_{all}$ & $t_{SCF}$ &$t_{MA}$\\
			\midrule
			single-mesh&-7.9178&0.0008 &6,029,109  & 6,029,109&18613.60 &13279.79  &5333.82\\\midrule
			multi-mesh&-7.9181&0.0007 &3,786,064 &3,665,015&14025.13 & 10245.13 &4262.25\\
			\bottomrule  
		\end{tabular}	
	\end{table}
	
\begin{figure}[!h]
		\centering
		\includegraphics[width=\linewidth]{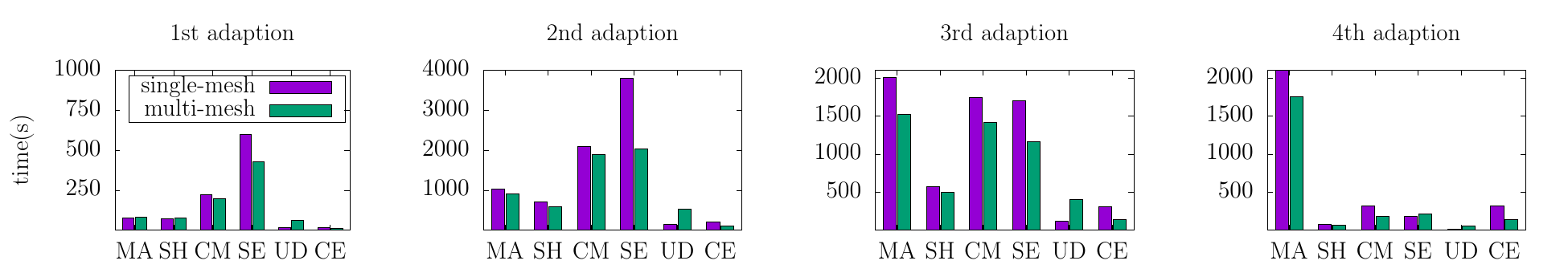}
		\caption{CPU time for single-mesh method and multi-mesh method for LiH molecule. Four mesh adaptations are needed to achieve chemical accuracy in both methods starting from the same initial setup. \label{fig:lihtime}}
	\end{figure}

	The detailed comparison regarding CPU time is illustrated in \Cref{fig:lihtime}. As expected, in the LiH example, the solution of the eigenvalue problem significantly contributes to the CPU time due to the increased number of required eigenvalues. Notably, the cost of the third mesh is higher than that of the final mesh. This is because, during the third mesh adaptation step, the mesh tends to stabilize with minimal changes, leading to fewer iteration steps in the final mesh, as shown in \Cref{fig:hehis}. Consequently, the CPU time on the third mesh becomes the most significant part contributing to the total CPU time. Moreover, since the number of Dofs in the multi-mesh method is smaller than that in the single-mesh method, the SCF time for the multi-mesh approach is also reduced. Overall, the multi-mesh method proves to be more efficient in terms of computational time than the single-mesh method in this case.
	
	\subsubsection{\ch{BeH2} molecule}
\begin{figure}[!h]
		\centering
		\includegraphics[width=\linewidth]{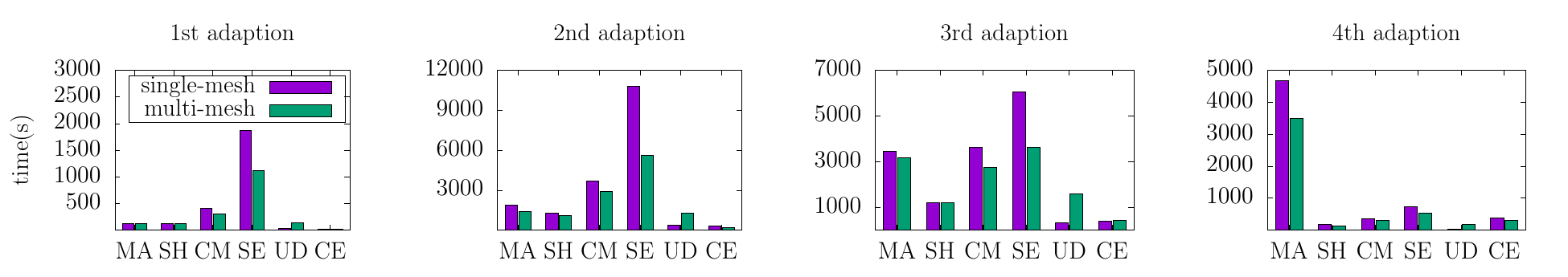}
		\caption{CPU time for single-mesh  method and multi-mesh method for \ch{BeH2} molecule. Four mesh adaptations are needed to achieve chemical accuracy in both methods starting from a same initial setup. \label{fig:beh2time}}
	\end{figure}

	The comparison between the multi-mesh method and the single-mesh method is also presented in the \ch{BeH2} example, as displayed in \Cref{tab:BeH2Time} and \Cref{fig:beh2time}. A similar conclusion to the previous examples on the accuracy and efficiency can be delivered from the results. In this case, given the same mesh adaption tolerance, the multi-mesh method succeeds in achieving the chemical accuracy, while the single-mesh method fails. This observation illustrates the superior accuracy through the multi-mesh method compared to the single-mesh method. 
	\begin{table}[!h]
		\caption{Comparison on \ch{BeH2} with respect to results and serial computational time. The referenced value is  $E_\mathrm{ref} = -15.6606$ Hartree. \label{tab:BeH2Time}}
		\centering
		\begin{tabular}{ rrrrrrrr}
			\toprule
			&$E_{tot}$ &$\Delta E/3$ &$N_{KS}$  &$N_{Har}$ & $t_{all}$& $t_{SCF}$ &$t_{MA}$\\
			\midrule
			single-mesh&-15.6573& 0.0011&7,338,638 & 7,338,638 &42445.26 & 32331.87
			&10113.40\\\midrule
			multi-mesh&-15.6580&0.0009 &4,852,528 &4,276,863&32133.92
			&23925.93
			&8207.99\\
			\bottomrule  
		\end{tabular}	
	\end{table}
	
	\subsection{Collection of results for more general electronic structures}
	
	In this subsection, we list the results of atoms and molecules using multi-mesh adaptive methods. The referenced values are generated from the state-of-art software \texttt{NWChem} using \emph{aug-cc-pv5z} basis sets except for the helium atom in which the \emph{aug-cc-pv6z} basis set is adopted. According to \Cref{tab:list}, we observe that chemical accuracy is achieved in the first four examples. However, as the number of atoms increases, the accuracy diminishes. This degradation can be attributed to the limitations of computational resources.  In such scenarios, we try to fully utilize our computational resources for simulations using both the single-mesh and multi-mesh adaptive methods, which can be implemented by dynamically adjusting the mesh adaption tolerance. Hence a comparative assessment between the two approaches can be delivered.
	\begin{table}[!h]
		\caption{List of examples. The referenced value is obtained from \texttt{NWChem}. $N_{s,\mathrm{KS}}$ represents the number of mesh grids for the single mesh method. $E_{s,\mathrm{KS}}$ stands for the energy obtained from the single-mesh method. The last row refers to the comparison with chemical accuracy. \label{tab:list}}
		\centering
		\begin{tabular}[tb]{lccccccc}
			\toprule
			molecule              & He       & LiH      &\ch{BeH2} &\ch{CH4}  &\ch{CH3OH}&\ch{CH3CH2OH} &\ch{C6H6} \\ \midrule
			$N_{s,\mathrm{KS}}$   &2,874,807 &6,029,109 &7,338,638 &8,121,252 &7,552,976 &7,720,132 &7,210,851  \\
			$N_{\mathrm{KS}}$     &1,262,599 &3,786,064 &4,852,528 &5,440,384 &6,226,903 &6,864,714 &6,036,918\\
			$N_{\mathrm{Har}}$    &2,185,920 &3,665,015 &4,276,863 &5,322,827 &4,063,275 &3,593,622 &4,798,747\\\midrule
			$E_\mathrm{ref}$ &-2.8348   &-7.9196   &-15.6606  &-40.1198  &-114.8503 &-153.8158 &-230.1916 \\
			$E_{s,\mathrm{KS}}$   &-2.8341   &-7.9178   &-15.6573  &-40.1105  &-114.8033 &-153.7418 &-229.9824  \\ 
			$E_{\mathrm{KS}}$     &-2.8340   &-7.9181   &-15.6580  &-40.1147  &-114.8205 &-153.7734 &-230.0663 \\ \midrule	
			$|\Delta E_{s,\mathrm{KS}}/E_\mathrm{ref}|$     &0.0002&0.0002&0.0002&0.0002&0.0004&0.0005&0.0009\\
			$|\Delta E_{\mathrm{KS}}/E_\mathrm{ref}|$       &0.0003&0.0002&0.0002&0.0001&0.0003&0.0003&0.0005\\
			$|\Delta E_{s,\mathrm{KS}}/n_{atom}|$&\textbf{0.0007}&\textbf{0.0009}&0.0011&0.0019&0.0078&0.0082&0.0174\\
			$|\Delta E_{\mathrm{KS}}/n_{atom}|$&\textbf{0.0008}&\textbf{0.0008}&\textbf{0.0009}&\textbf{0.0010}&0.0050&0.0047&0.0104\\
			\bottomrule
		\end{tabular}
	\end{table}
	
	\begin{figure}[h]
		\centering
		\includegraphics[width=0.18\linewidth,frame]{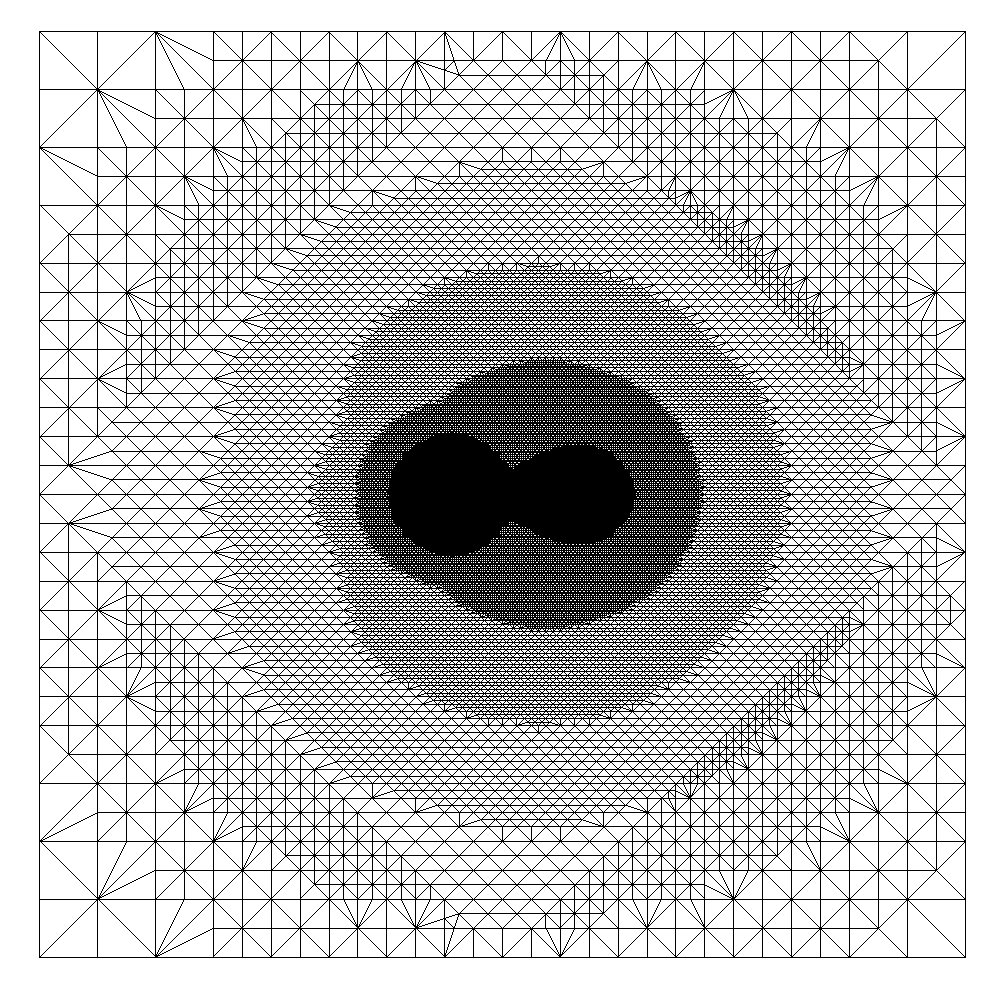}~
		\includegraphics[width=0.18\linewidth,frame]{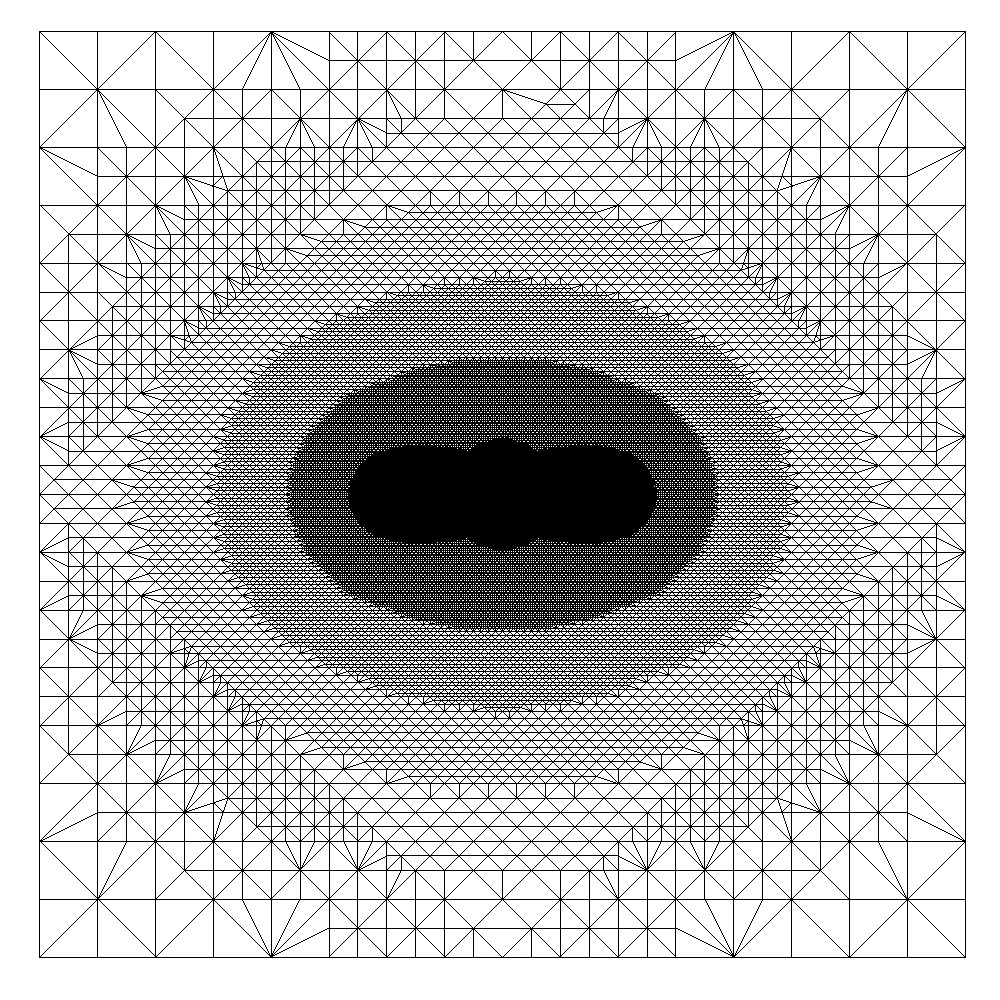}~
		\includegraphics[width=0.18\linewidth,frame]{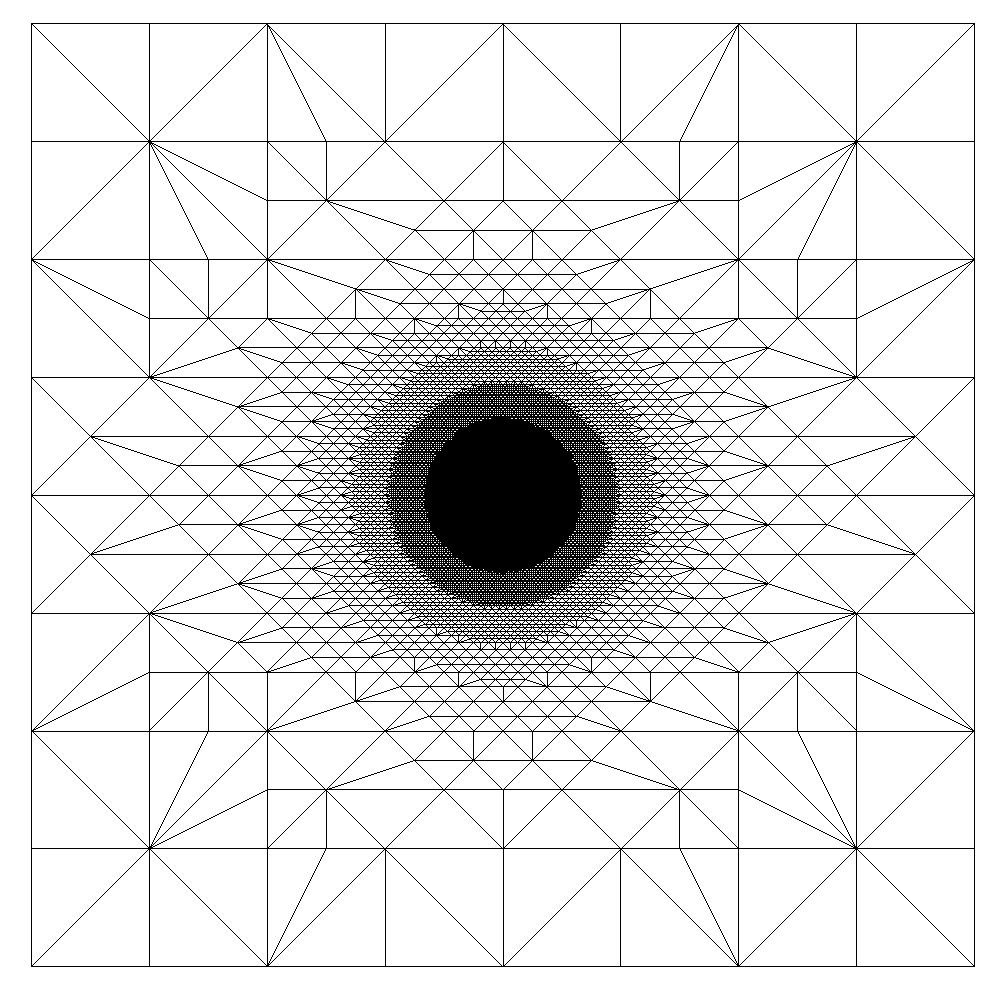}~
		\includegraphics[width=0.18\linewidth,frame]{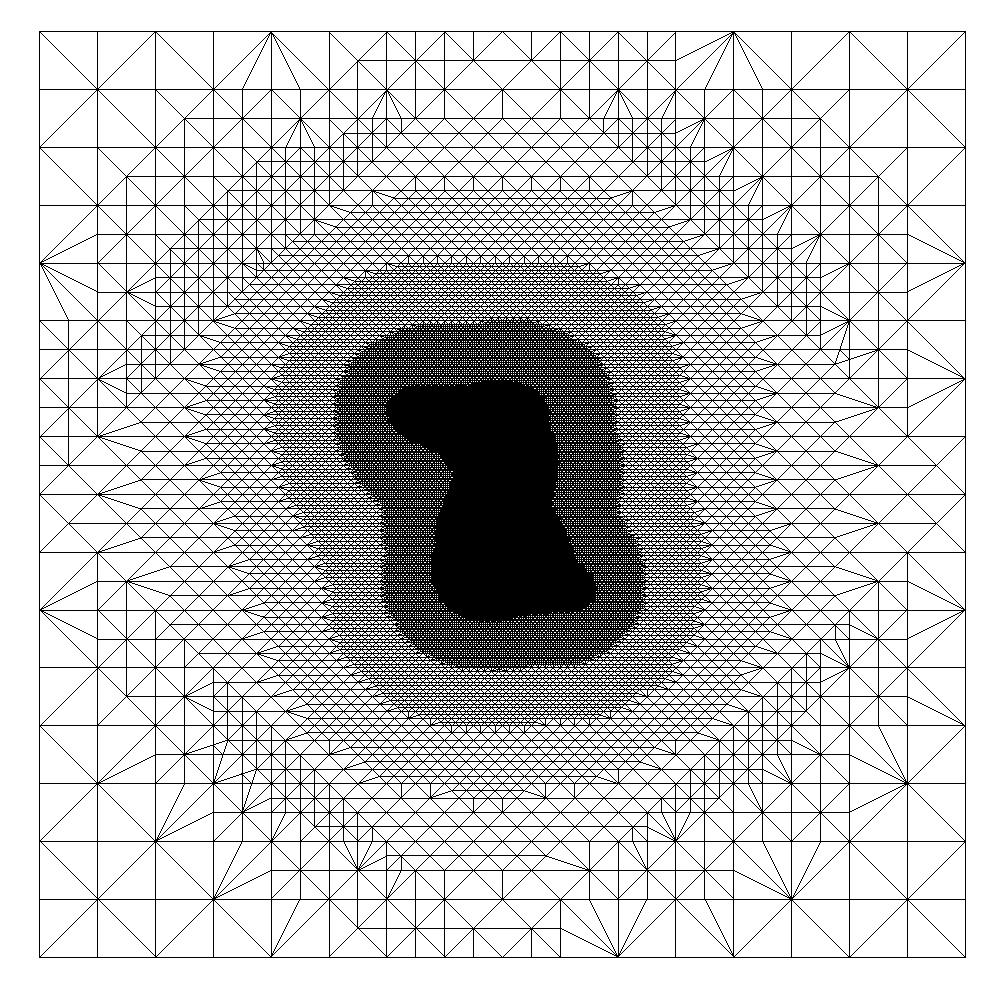}~
		\includegraphics[width=0.18\linewidth,frame]{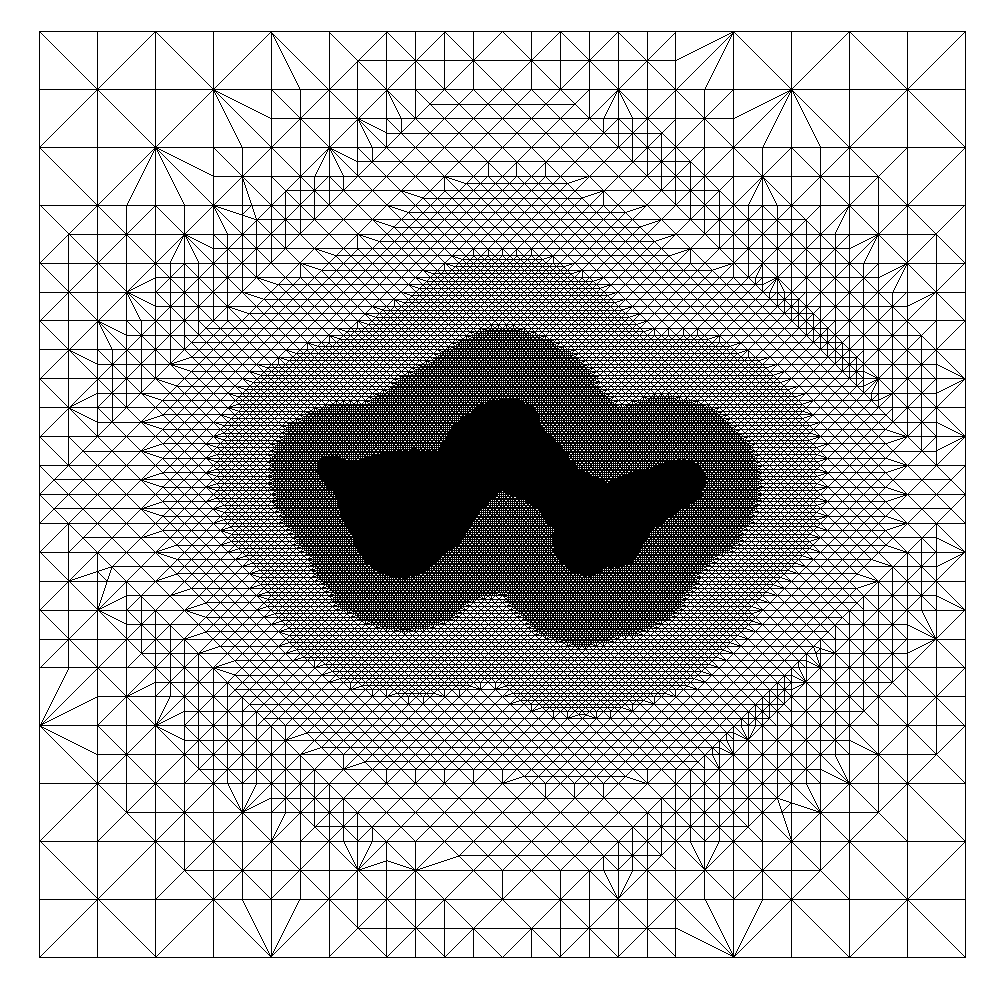}	\\
		\includegraphics[width=0.18\linewidth,frame]{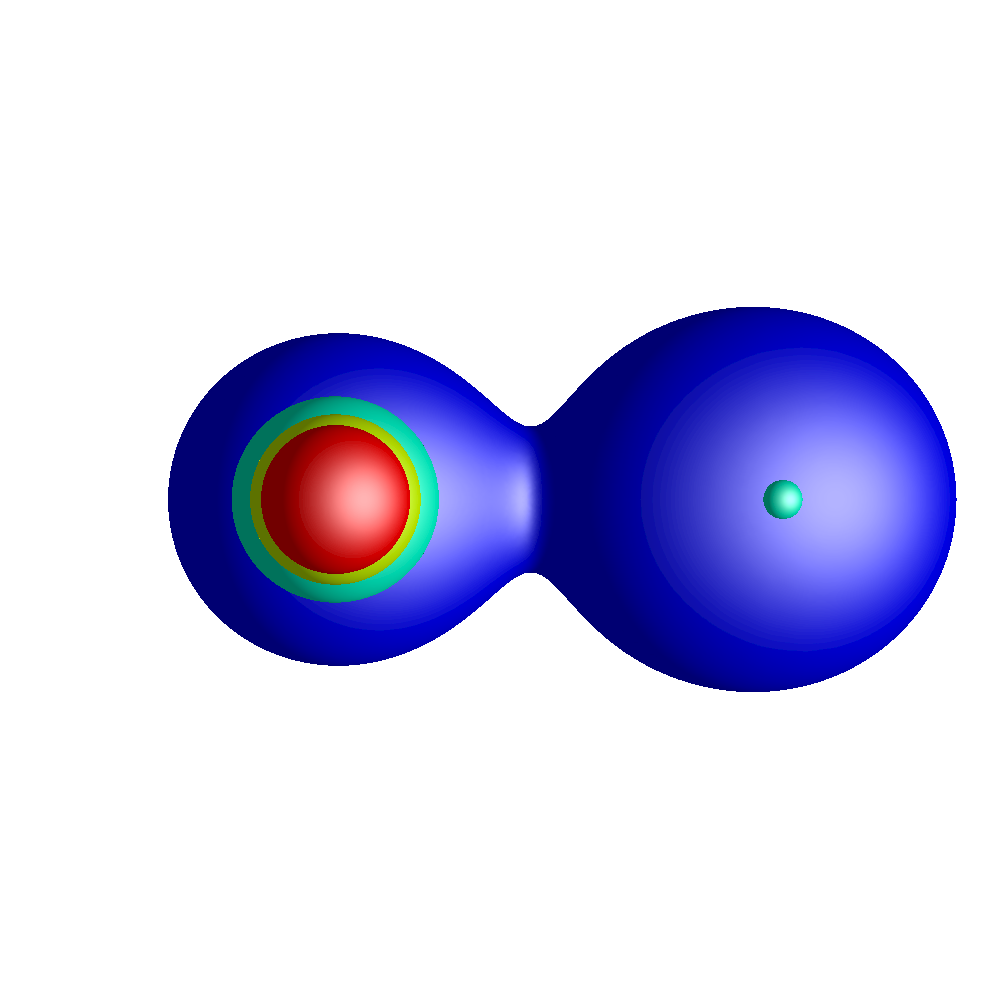}~
		\includegraphics[width=0.18\linewidth,frame]{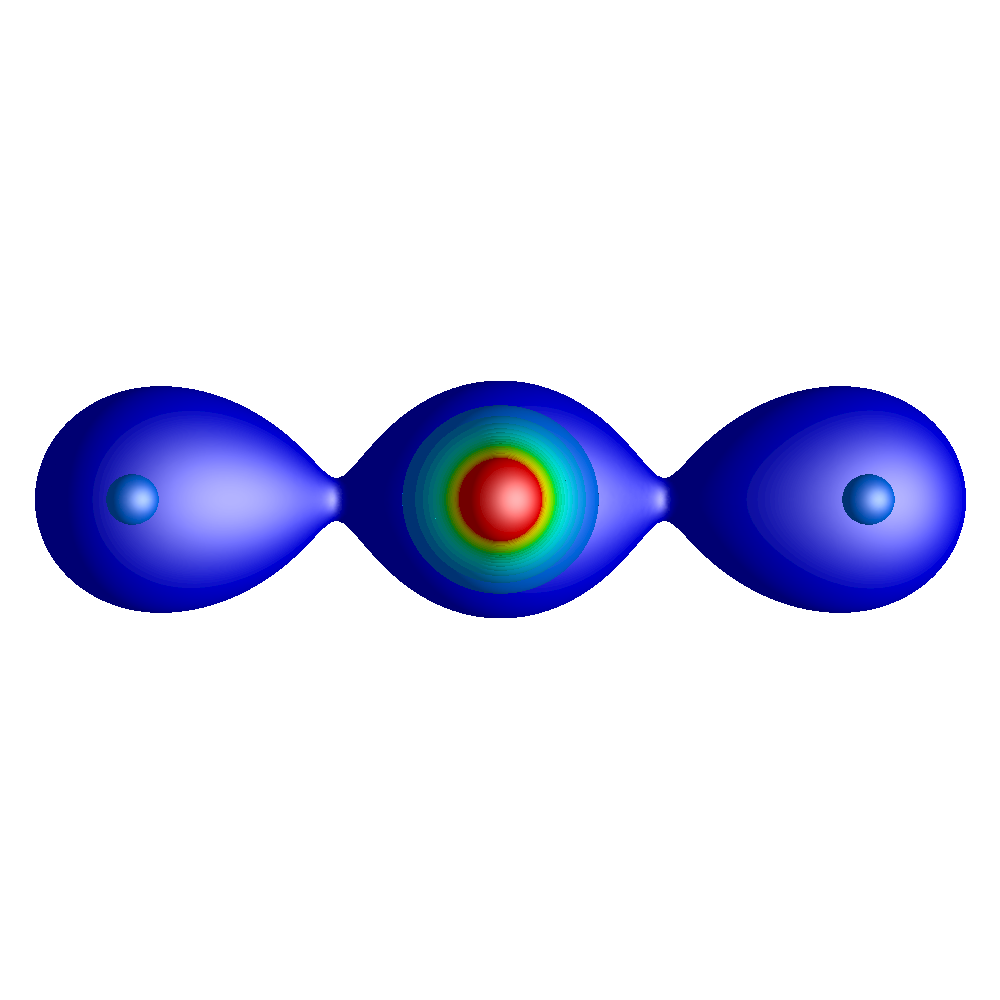}~
		\includegraphics[width=0.18\linewidth,frame]{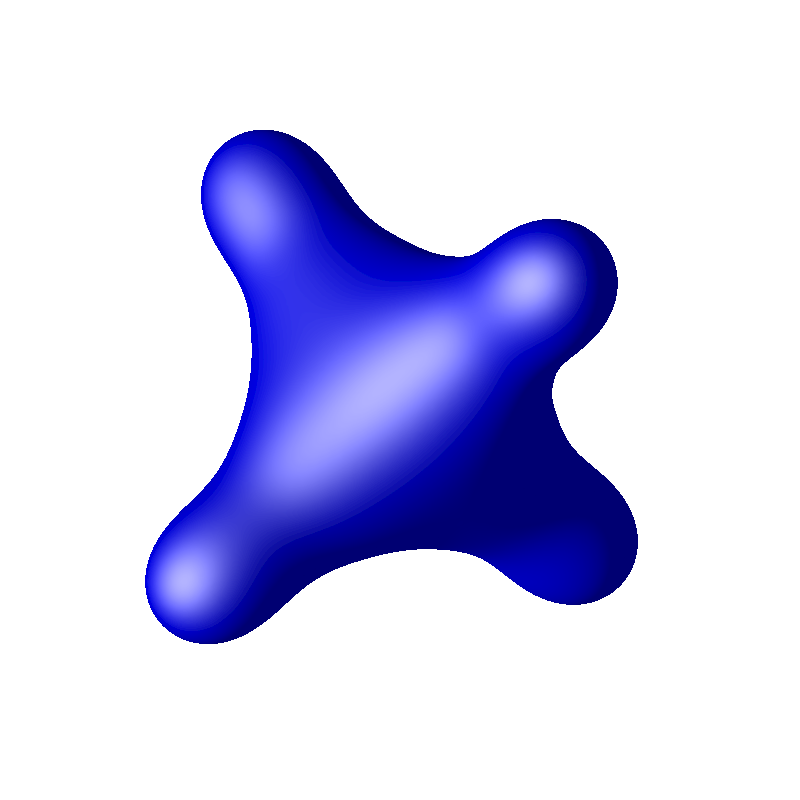}~
		\includegraphics[width=0.18\linewidth,frame]{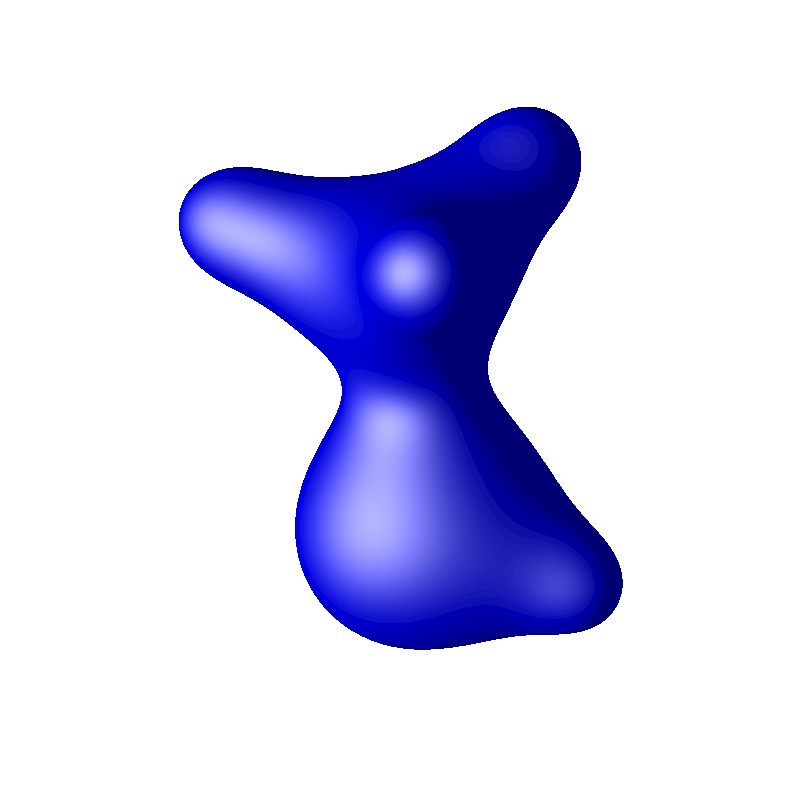}~
		\includegraphics[width=0.18\linewidth,frame]{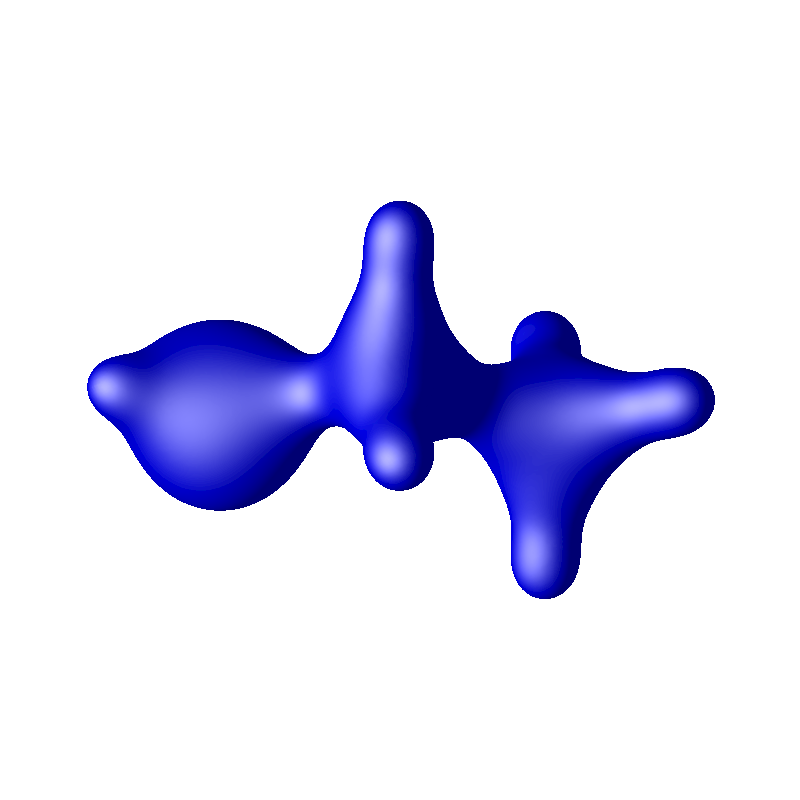}		
		\caption{Sliced KS meshes (top) and isosurfaces (bottom) for the following molecules: \ch{LiH}, \ch{BeH2}, \ch{CH4}, \ch{CH3OH}, and \ch{CH3CH2OH}. Contours are depicted for the two-dimensional structures (\ch{LiH} and \ch{BeH2}). The computational domain is set as $[-10,10]^3$, with the exception of the \ch{CH4} example, for which it is extended to $[-20,20]^3$. \label{fig:list}}	
	\end{figure}
	
	The results are summarized in \Cref{tab:list} and we display several meshes and isosurfaces for these examples in \Cref{fig:list}. From the table and the figure, we can see that 
	\begin{itemize}
		\item For the last four molecules, besides a similar observation for the CPU time in Figures \ref{fig:hetime}, \ref{fig:lihtime}, and \ref{fig:beh2time}, another observation can be made, that with comparable computational resource, our multi-mesh method can generally produce a much better result. It means that the multi-mesh method would be an answer to the question of how to generate the most accurate result, by fully using the given computational resource.
		\item When maximizing our computational resources, the multi-mesh approach yields more precise results. To be specific, the multi-mesh method attains chemical accuracy for the first four examples, as indicated in the last row of \Cref{tab:list}. In contrast, the single-mesh method only achieves chemical accuracy in the first two examples.
		
		\item In the multi-mesh method, the number of mesh grids for the Hartree mesh exceeds the number for the KS mesh only in the case of helium. As the system size scales up, the number of Hartree mesh grids becomes progressively less than the number of KS mesh grids. Consequently, the computational time needed for the Hartree potential becomes less significant as the system size increases. There is an exception in the cases of \ch{CH4} and \ch{C6H6}, which arises because we opted for a larger computational domain, specifically, $[-20,20]^3$.
	\end{itemize}

	\section{Conclusion}
	In this paper, we introduce a multi-mesh adaptive finite element framework for Kohn-Sham density functional theory, aiming to achieve chemical accuracy. We investigate the impact of the Hartree potential approximation on total energy, finding that chemical accuracy cannot be attained without a well-approximated Hartree potential. While the single-mesh adaptive method, considering the Hartree potential, achieves chemical accuracy, it comes with significant computational costs. To address this, we propose the multi-mesh adaptive method, which offers a more efficient route to achieving chemical accuracy.
	
	We demonstrate the effectiveness and accuracy of the multi-mesh adaptive method through various numerical examples. However, we observe that even for the helium atom, achieving chemical accuracy demands a substantial number of mesh grids, mainly due to the use of linear Lagrange finite elements. One strategy is to adopt higher-order elements under the multi-mesh framework. Additionally, we consider future work on conducting numerical analysis for the multi-mesh adaptive method. Moreover, we recognize the need for accelerating the multi-mesh adaptive approach, particularly through parallel implementation and the development of more effective error indicators for mesh adaptation methods. These enhancements will further refine the efficiency and accuracy of the multi-mesh adaptive framework.

	\section*{Acknowledgement}
	The work of Y. Kuang was supported in part by the National Natural Science Foundation of China (Grant No.12201130) and the Guangzhou Municipal Science and Technology Bureau (Grant No.2023A04J1321).  The work of Y. Shen was supported by Hunan Key Laboratory for Computation and Simulation in Science and Engineering (Grand No. LCSSE202307). The work of G. Hu was supported by the National Natural Science Foundation of China (Grant Nos.11922120 and 11871489), the FDCT of MacaoSAR (Grant No.0082/2020/A2), the MYRG of the University of Macau (Grant No. MYRG2020-00265-FST), and the Guangdong-Hong Kong-Macao Joint Laboratory for Data Driven Fluid Mechanics and Engineering Applications (Grant No. 2020B1212030001).

	\bibliography{biblio}
	\bibliographystyle{plain}

\end{document}